\documentclass[proof,pdftex]{pasj00}
\draft

\begin{document}
\SetRunningHead{M\"uller \& Hasegawa}{Itokawa revisited}

\title{(25143)~Itokawa: The Power of Radiometric Techniques for the
     Interpretation of Remote Thermal Observations in the Light of
     the Hayabusa Rendezvous Results\thanks{Based
     on observations with AKARI, a JAXA
     project with the participation of ESA.}}

\author{Thomas G. \textsc{M\"uller}}
\affil{%
    Max-Planck-Institut f\"ur extraterrestrische Physik,
    Giessenbachstra{\ss}e, 85748 Garching, Germany
    }
\email{tmueller@mpe.mpg.de}

\author{Sunao \textsc{Hasegawa}}
\affil{%
    Institute of Space and Astronautical Science,
    Japan Aerospace Exploration Agency,
    3-1-1 Yoshinodai, Chuo-ku, Sagamihara 252-5210
    }
\email{hasehase@isas.jaxa.jp}

\and 

\author{Fumihiko \textsc{Usui}}
\affil{%
    Department of Astronomy, Graduate School of Science,
    The University of Tokyo, 7-3-1 Hongo, Bunkyo-ku,
    Tokyo 113-0033
    }
\email{usui@astron.s.u-tokyo.ac.jp}


\maketitle

\begin{abstract}
          The near-Earth asteroid (25143)~Itokawa 
          was characterised in great detail by the Japanese
          Hayabusa mission. We revisited the available thermal observations
          in the light of the true asteroid properties
          with the goal to evaluate the possibilities and limitations
          of thermal model techniques. In total, we used 25
          published ground-based mid-infrared photometric observations
          and 5 so far unpublished measurements from the Japanese infrared 
          astronomical satellite AKARI in combination with improved H-G values
          (absolute magnitude and slope parameter).
          Our thermophysical model (TPM) approach
          allowed us to determine correctly the sense of rotation,
          to estimate the thermal inertia and to derive robust effective
          size and albedo values by only using a simple spherical shape model.
          A more complex shape model, derived from light-curve
          inversion techniques, improved the quality of the predictions
          considerably and made
          the interpretation of thermal light-curve possible.
          The radiometrically derived effective diameter value agrees
          within 2\% of the true Itokawa size value.
          The combination of our TPM and the final (25143)~Itokawa
          in-situ shape model was then used as a benchmark for deriving
          and testing
          radiometric solutions. The consolidated value for the surface-averaged
          thermal inertia is $\Gamma$\,=\,700 $\pm$ 200\,J\,m$^{-2}$\,s$^{-0.5}$\,K$^{-1}$.
          We found that even the high resolution shape models still require
          additional small-scale roughness in order to explain the disk-integrated 
	  infrared measurements.
          Our description of the thermal effects as a function of wavelengths,
          phase angle,
          and rotational phase
          facilitates the planning of crucial thermal
          observations for sophisticated characterization of small bodies, including
          other potentially hazardous asteroids. Our analysis shows the
          power of radiometric techniques to derive the size, albedo,
          thermal inertia, and also spin-axis orientation from small sets
          of measurements at thermal infrared wavelengths.
\end{abstract}


\section{Introduction}

  The Near-Earth asteroid (NEA) (25143)~Itokawa (1998\,SF36) is one
  of the best studied asteroids in our Solar System. It was
  the sample return target of the Japanese Hayabusa (MUSES-C) mission.
  The spacecraft was in close proximity to the asteroid
  from September through early December 2005. As a result of the
  encounter, the asteroid has been characterised in great detail:
  (25143)~Itokawa is an irregularly formed body consisting of a loose pile
  of rubble rather than a solid monolithic asteroid (\cite{fujiwara06, saito06, abe06b}).
  Its appearance is boomerang-shaped and composed of two distinct parts
  with faceted regions and a concave ring structure in-between
  (\cite{demura06}). Recently, the detection of YORP spin-up revealed that the two
  distinct parts of Itokawa have different densities and are likely to be
  two merged asteroids (\cite{lowry14}).
  Its effective diameter (of an equal volume sphere)
  is 327.5$\pm$5.5\,m (volume of (1.840$\pm$0.092)$\times$10$^{7}$\,m$^{3}$;
  \cite{fujiwara06}).
  This compares very well with the pre-encounter size prediction
  obtained via radiometric techniques by
  \citet{mueller05b} (M05 hereafter) of D$_{\mathrm{eff}}$=320$\pm$30\,m.
  The radar size prediction is about 16\% too high
  (\cite{ostro04, ostro05}). The mass is estimated as
  (3.58$\pm$0.18)$\times$10$^{10}$\,kg, implying a bulk density of
  (1.95$\pm$0.14)\,g cm$^{-3}$ (\cite{abe06b}).
  The retrograde pole orientation
  in ecliptic coordinates is
  ($\lambda_{pole}$, $\beta_{pole}$) = (128.5$^{\circ}$, -89.66$^{\circ}$),
  with a 3.9$^{\circ}$ margin of error (\cite{demura06}).
  Itokawa is classified as an S\,IV-type asteroid via ground-based
  near-infrared (NIR) spectroscopy (\cite{binzel01}), a type common in the inner
  portion of the asteroid belt. These measurements at mineralogically
  diagnostic wavelength show similarities to ordinary chondrites
  and/or primitive achondrite meteorites. Hayabusa confirmed the S-class
  asteroid characteristics and revealed an olivine rich mineral assemblage
  of the surface, similar to LL5 or LL6 chondrites (\cite{abe06a, okada06}).
  The most recent geometric albedo estimate of $p_V\,=\,$0.29\,$\pm$\,0.02 comes
  from ground-based visual photometry combined with Hayabusa-derived
  size information (\cite{bernardi09}).
  The surface is dominated by regions with brecciated rocks and
  regions with a coarse-grain-filled surface with thermal inertias
  between that of monolithic rocks ($\Gamma \sim4000$\,J\,m$^{-2}$\,s$^{-0.5}$\,K$^{-1}$)
  and powdery surface like lunar regolith ($\Gamma \sim40$\,J\,m$^{-2}$\,s$^{-0.5}$\,K$^{-1}$)
  (\cite{yano06, noguchi10}). M05 used a sample of remote, disk-integrated 
  thermal measurements to derive the average thermal inertia of Itokawa's top
  surface layer. They found a thermal inertia value of roughly 750\,J\,m$^{-2}$\,s$^{-0.5}$\,K$^{-1}$.
  A study by \citet{mueller07} found a very similar value of 700\,J\,m$^{-2}$\,s$^{-0.5}$\,K$^{-1}$.
  \citet{gundlach13} combined the thermal inertia value with Itokawa's known size and low gravitational
  acceleration on the surface to determine a mean surface particle radius of 21$^{+3}_{-14}$\,mm
  which is in nice agreement with in-situ observations presented by \citet{yano06} and
  \citet{kitazato08}.
  One aspect of our work here was to test the derived (pre-Hayabusa) thermal properties
  (mainly the object's thermal inertia) in the light of the in-situ results.
  
  We re-visit the available remote, disk-integrated thermal data, comprising
  ground-based observations in standard N- and Q-band filters with IRTF/MIRSI and
  ESO/TIMMI2, and from AKARI, the Japanese infrared astronomical satellite
  (\cite{murakami07}). The goal of our study is to describe the possibilities
  and limitations of radiometric methods when using remote, disk-integrated
  data, as available or easily obtainable for most of the minor bodies.
  The derived values are then directly compared to the in-situ results
  from the Hayabusa-mission. By comparing the results from the radiometric
  techniques with the in-situ results, we validate model techniques
  and provide observing strategies for future applications to
  other targets, including potentially hazardous asteroids (PHAs).

  Section~\ref{cha:obs} gives an overview of the existing thermal
  observations and describes the so far unpublished
  observations by AKARI.
  In Section~\ref{cha:tpm} we describe the different TPM applications
  and optimisation processes which we applied to the full dataset
  of remote thermal observations and present the derived results:
  Section~\ref{sec:tpm} describes briefly the thermophysical model (TPM)
  and the range of possible input parameters. In a first analysis step
  we use a spherical shape model with a range of pro- and retrograde
  spin-vectors (Section~\ref{sec:sphere}).
  In the second step (Section~\ref{sec:trishape}) we added the shape and
  spin-vector information derived from light-curve inversion techniques.
  In the third step (Section~\ref{sec:insitu}) we used the true shape
  model and spin-vector solution as provided by the Hayabusa mission.
  The flux predictions from the best TPM solution are then compared
  with the available observations. 
  In Section~\ref{cha:predictions} we inter-compare different shape solutions
  with respect to thermal light-curves and predict the behaviour of the
  spectral energy distribution (SED), the thermal light-curve amplitude, shape,
  and the thermal beaming effect.
  In Section~\ref{cha:discussion} we discuss the
  potential and the limitation of the radiometric methods, using the
  various levels of information.
 The summary and transfer of applications
 to other targets is given in Section~\ref{cha:conclusions}.

  
\section{Thermal Observations and Input Data}
\label{cha:obs} 

  \begin{table*}
  \begin{center}
    \caption{Summary of mid-IR observing sets for asteroid (25143)~Itokawa. }
             \label{tbl:obslog}
    \begin{tabular}{lllllll}
      \hline
      \hline
      \noalign{\smallskip}
           & \multicolumn{1}{c}{Time\footnotemark[$\dag$]} & Filter & $r$\footnotemark[$\ddag$]    & $\Delta$\footnotemark[$\S$] & $\alpha$\footnotemark[$\|$]     & \\
      No\footnotemark[$*$]   & \multicolumn{1}{c}{[UT]} & Band   & [AU] & [AU]     & [$^{\circ}$] & Remarks \\
      \noalign{\smallskip}
      \hline
      \noalign{\smallskip}
      21 & 2004/Jul/10~11:45 & N11.7 & 1.060399 & 0.049983 &  +28.32 & IRTF/MIRSI (\cite{mueller05a})\footnotemark[$\#$] \\
      22 & 2004/Jul/10~11:48 & N11.7 & 1.060407 & 0.049990 &  +28.31 &  IRTF/MIRSI (\cite{mueller05a})\footnotemark[$\#$] \\
      23 & 2004/Jul/10~13:32 & N11.7 & 1.060673 & 0.050243 &  +28.18 &  IRTF/MIRSI (\cite{mueller05a})\footnotemark[$\#$] \\
      24 & 2004/Jul/10~13:41 & N9.8  & 1.060696 & 0.050266 &  +28.17 &  IRTF/MIRSI (\cite{mueller05a})\footnotemark[$\#$] \\
      25 & 2004/Jul/10~13:51 & N9.8  & 1.060721 & 0.050290 &  +28.16 & IRTF/MIRSI (\cite{mueller05a})\footnotemark[$\#$] \\
      \noalign{\smallskip}
      26 & 2007/Jul/26~11:29 &    N4 & 1.053777 & 0.281244 &  -73.49 & AKARI (this work)\footnotemark[$**$] \\
      27 & 2007/Jul/26~11:29 &    S7 & 1.053777 & 0.281244 &  -73.49 & AKARI (this work)\footnotemark[$**$] \\
      28 & 2007/Jul/26~11:28 &   S11 & 1.053774 & 0.281244 &  -73.49 & AKARI (this work)\footnotemark[$**$] \\
      29 & 2007/Jul/26~13:09 &   L18 & 1.054027 & 0.281282 &  -73.43 & AKARI (this work)\footnotemark[$**$] \\
      30 & 2007/Jul/26~13:12 &   L24 & 1.054035 & 0.281283 &  -73.43 & AKARI (this work)\footnotemark[$**$] \\
      \noalign{\smallskip}
    \noalign{\smallskip}
    \hline
    \multicolumn{7}{l}{\hbox{\parbox{160mm}{\footnotesize
          \par\noindent
              {Notes.
               \footnotemark[$*$]{Observations with running numbers 1-20 are listed
                                 in table~1 in M05.}
	       \footnotemark[$\dag$]{The times are mid observing times in the observer's time frame. }
               \footnotemark[$\ddag$]{The heliocentric distance. }
               \footnotemark[$\S$]{The observer-centric distance. }
               \footnotemark[$\|$]{The phase angles, negative before opposition and positive after. }
               \footnotemark[$\#$]{The observations in \citet{mueller05a}
                                    have been shifted to the observer's
                                    time frame by adding 25\,s to the light-time
                                    corrected times given in the publication.}
               \footnotemark[$**$]{The geometry is given by the geocentric calculation.}
               }
        }\hss}}
    \end{tabular}
  \end{center}
  \end{table*}

  We combine five previously published mid-infrared observations
  by \citet{mueller05a} with 20 observations
  by M05 and five dedicated AKARI observations. The M05 data (table~1 \& 3 in M05)
  have running indices from 1 to 20. The additional data presented here are
  labeled 21-25 and 26-30 respectively (table~\ref{tbl:obslog} and
  table~\ref{tbl:obsres}).

\subsection{IRTF/MIRSI observations}

  \citet{mueller05a} presented a set of five N-band observations
  which we included in our calculations. For the entries in table~\ref{tbl:obslog} and
  table~\ref{tbl:obsres} (numbers 21-25) we used the monochromatic,
  colour-corrected fluxes (but now in Jansky-units) and calculated the
  true observing times (\cite{mueller05a}
  gave times which were corrected for 1-way light-time, i.e., in the asteroid
  time frame). \citet{mueller07} mentioned
  that they had taken the observations at a relatively high level of atmospheric
  humidity.
  In addition, all observations were taken at air-masses larger than 2 due to technical
  problems at meridian transit.

\subsection{AKARI observations}

  Asteroid (25143)~Itokawa was observed on July 27, 2007 by the
  NIR, MIR-S, and MIR-L channels on the infrared camera IRC
  (\cite{onaka07}) on-board AKARI.
  During one pointed observation, all three IRC channels obtained images
  simultaneously, covering different wavelength ranges. The NIR and MIR-S
  channels share the same field of view, while the MIR-L channel observes a
  region which is $\sim${20$^{\prime}$} away from the field centre of
  the NIR and MIR-S channels. In total, two pointed observations on Itokawa
  were carried out to obtain data in all three channels.
  The Astronomical Observation Template (AOT) IRC02 for dual-filter photometry
  (see \cite{onaka07} for details) was used.
  As a result, observations for Itokawa with the NIR, MIR-S, and MIR-L were
  performed in the filters N3 (reference wavelength of 3.2\,$\mu$m, but not
  used here for our thermal analysis), N4 (4.1\,$\mu$m),
  S7 (7.0\,$\mu$m), S11 (11.0\,$\mu$m), L15 (15.0\,$\mu$m)
  and L24 (24.0\,$\mu$m) with effective bandwidths of 0.9, 1.5, 1.8, 4.1, 6.0
  and 5.3\,$\mu$m, respectively. The projected area of the NIR channels was
  about 10.0$^{\prime}$ $\times$ 9.5$^{\prime}$ which corresponds to an
  angular resolution of 1.5$^{\prime \prime}$/pixel. The MIR-S channels
  of the IRC have pixel sizes of about 2.3$^{\prime \prime}$/pixel,
  giving a field of view about 10.0$^{\prime}$ $\times$ 9.1$^{\prime}$.
  The MIR-L channel was used with a image scale of 2.4$^{\prime \prime}$/pixel,
  giving a 10.2$^{\prime}$ $\times$ 10.3$^{\prime}$ sky field.
  For the data processing the IRC imaging data pipeline\footnote{AKARI IRC Data
  Users Manual ver.1.3, {\tt http://www.ir.isas.jaxa.jp/ASTRO-F/Observation/}}
  was used.
  The AKARI telescope was not able to track moving objects such as comets and asteroids.
  Therefore, a centroid determination in combination with a standard shift-and-add technique
  was performed, followed by median processing to obtain better photometric accuracy.
  Aperture photometry on IRC images was carried out using the APPHOT task of IRAF
  thorough circular aperture radii of 10.0 (in the NIR channel) and 7.5 (in the MIR-S and MIR-L channels)
  pixels, which are also used for the standard star flux calibration.
  The resulting astronomical data units were converted to the calibrated flux densities by using the
  IRC flux calibration constants in the {\it Revisions of the IRC conversion
  factors}\footnote{\tt http://www.ir.isas.jaxa.jp/ASTRO-F/Observation/\-DataReduction/IRC/ConversionFactor\_071220.html}.
  Colour differences between calibration stars and Itokawa were not negligible due to the wide
  bandwidths of the IRC.
  Colour correction factors were obtained using both predicted thermal flux of Itokawa and
  the relative spectral response functions for IRC.
  Colour correction fluxes of Itokawa were obtained by dividing the quoted fluxes by
  1.453 in N4 band, 1.020 in S7 band, 0.956 in S11 band, 0.960 in L18 band,
  and 1.079 in L24 band.
  The observational results are summarised in table~\ref{tbl:obsres} (numbers 26-30)
  and further details about AKARI asteroid observations and catalogued data
  are given in \citet{usui11, hasegawa13}.

  \begin{table*}
    \begin{center}
    \caption{Summary of the available thermal infrared observations of asteroid (25143)~Itokawa.
             \label{tbl:obsres}} 
    \begin{tabular}{llrlll}
      \hline
      \hline
      \noalign{\smallskip}
           &        & $\lambda_c$ & FD   & $\sigma_{\mathrm{err}}$  &  \\
      No\footnotemark[$*$]   & Filter & [$\mu$m]    & [Jy] & [Jy]            & Remarks \\
      \noalign{\smallskip}
      \hline
      \noalign{\smallskip}
      21 & N11.7 & 11.7  & 0.762 & 0.100 & IRTF/MIRSI (\cite{mueller05a})\footnotemark[$\dag$] \\
      22 & N11.7 & 11.7  & 0.721 & 0.091 & IRTF/MIRSI (\cite{mueller05a})\footnotemark[$\dag$] \\
      23 & N11.7 & 11.7  & 0.913 & 0.114 & IRTF/MIRSI (\cite{mueller05a})\footnotemark[$\dag$] \\
      24 & N9.8  &  9.8  & 0.791 & 0.125 & IRTF/MIRSI (\cite{mueller05a})\footnotemark[$\dag$] \\
      25 & N9.8  &  9.8  & 0.570 & 0.122 & IRTF/MIRSI (\cite{mueller05a})\footnotemark[$\dag$] \\
      \noalign{\smallskip}
      26 &   N4  &  4.1   &  0.00032  &  0.00025 & AKARI (this work)\\
      27 &   S7  &  7.0   &  0.00469  &  0.00028 & AKARI (this work)\\
      28 &  S11  & 11.0   &  0.01422  &  0.00053 & AKARI (this work)\\
      29 &  L15  & 15.0   &  0.02137  &  0.00079 & AKARI (this work)\\
      30 &  L24  & 24.0   &  0.01947  &  0.00120 & AKARI (this work)\\
      \noalign{\smallskip}
    \noalign{\smallskip}
    \hline
    \multicolumn{6}{l}{\hbox{\parbox{120mm}{\footnotesize
          \par\noindent
          Notes. 
          \footnotemark[$*$]{Observations with running numbers 1-20 are listed
                              in table~1 in M05.}
           \footnotemark[$\dag$]{The flux densities in \citet{mueller05a}
                              have been converted to Jansky.}
        }\hss}}

    \end{tabular}
    \end{center}
  \end{table*}


\section{Thermophysical Modelling}
\label{cha:tpm}

\subsection{Description of the Thermophysical Model}
\label{sec:tpm}

  We applied the radiometric technique as described in M05. Via a $\chi^2$-process,
  with diameter and thermal inertia as free parameters\footnote{We also solve for
  the geometric albedo, but is not considered as a free parameter since it is
  tightly connected to the H-magnitude via the size information:
  p$_V$ = 10$^{(2\cdot\,log_{10}(S_0) - 2\cdot\,log_{10}(D_{eff}) - 0.4\cdot\,H_V)}$,
  with the Solar constant S$_0$ = 1366\,W\,m$^{-2}$.},
  we searched for the best solution to match all thermal observations listed in Sect.~\ref{cha:obs}
  simultaneously: $\chi^2 = 1/(N - \nu) \Sigma ((obs - mod)^2)$, with $\nu$ being
  the number of free parameters (here $\nu$=2, with size and thermal inertia as free parameters).
  The detailed steps are described in \citet{mueller11}.
  The TPM is detailed by \citet{lagerros96, lagerros97, lagerros98a, lagerros98b, harris02}. 
  It places
  the asteroid at the true illumination and observing geometry. For each
  surface element the solar insolation is taken into account and the amount of 
  reflected light and thermal emission are calculated, controlled by
  the albedo, the H-G values, the surface roughness (parameterised by
  $\rho$, the r.m.s.\ of the surface slopes and $f$,
  the fraction of the surface covered by craters) and the thermal
  inertia $\Gamma$. For the temperature calculation the
  one-dimensional vertical heat conduction (controlled by the thermal 
  inertia\footnote{The thermal inertia $\Gamma$ is defined
  as $\sqrt{\kappa \rho c}$, where $\kappa$ is the thermal conductivity,
  $\rho$ is the density, and $c$ is the heat capacity.} $\Gamma$)
  into the surface is taken into into account.
  The treatment of heat conduction inside
  the spherical section craters is approximated by using the brightness temperature
  relations as a function of the thermal parameter\footnote{The thermal
  parameter $\Theta$ is defined as ($\Gamma$ $\sqrt{\omega}$) / ($\epsilon \sigma T_{ss}^3$),
  where $\Gamma$ is the thermal inertia, $\omega$ is the angular velocity
  of rotation and $T_{ss}$ is the sub-solar temperature.}
  $\Theta$ for a flat surface (\cite{lagerros98a}).
  In this way it is possible to separate the beaming from the heat conduction
  which is relevant for computation speed reasons.
  A summary of the influences of the thermal parameters as a function
  of wavelength and as a function of phase angle is given in 
  \citet{mueller02a}. The technique to determine thermal properties
  from a set of thermal observations was already successfully applied for
  large main-belt asteroids by e.g., \citet{spencer89, mueller98, mueller02, mueller99, orourke12}
  and a range of near-Earth asteroids (e.g., \cite{mueller11, mueller12, mueller13}).
  
  The general TPM input parameters and parameter 
  ranges are listed in table~\ref{tbl:tpm_input}. The first three parameters
  show the physically meaningful range for thermal properties
  (see e.g., \cite{lagerros98b}). The constant emissivity is a
  standard value used in radiometric techniques when applied to 
  mid-infrared data (e.g., \cite{lebofsky86}). The last three values are derived from visual
  photometric measurements.

  \begin{table*}
    \begin{center}
    \caption{Summary of general TPM input parameters and applied variations.
             \label{tbl:tpm_input}}
    \begin{tabular}{lcll}
      \hline
      \hline
      \noalign{\smallskip}
               & Range & Units/Remarks & M05 value \\
      \noalign{\smallskip}
      \hline
      \noalign{\smallskip}
    $\Gamma$            & 0...2500                  & [J\,m$^{-2}$\,s$^{-0.5}$\,K$^{-1}$] & 750 \\
                        &                           & thermal inertia & \\
    $\rho$              & 0.1...0.9                 & rms.\ of surface slopes & 0.7 \\
    $f$                 & 0.4...0.9                 & fraction of surface & 0.6 \\
                        &                           & covered by craters &  \\
    $\epsilon$          & 0.9                       & $\lambda$-independent emissivity & 0.9 \\
      \noalign{\smallskip}
    $H_{\rm{V}}$  & 19.40\,$^{+0.10}_{-0.09}$       & [mag] & 19.9 \\
                        &                           & \citet{bernardi09} & \\
      \noalign{\smallskip}
    G                   & 0.21$^{+0.07}_{-0.06}$ & \citet{bernardi09} & 0.21 \\
      \noalign{\smallskip}
    P$_{\mathrm{sid}}$           & 12.13237                  & [h] &  12.13237\\
                        & $\pm$0.00008            & \authorcite{kaasalainen03} (\yearcite{kaasalainen03}, \yearcite{kaasalainen05}) & \\
     \noalign{\smallskip}
     \noalign{\smallskip}
     \hline
    \end{tabular}
    \end{center}
  \end{table*}

\subsection{Using a spherical shape model}
\label{sec:sphere}

  \begin{figure}
    \begin{center}
      \includegraphics[width=80mm]{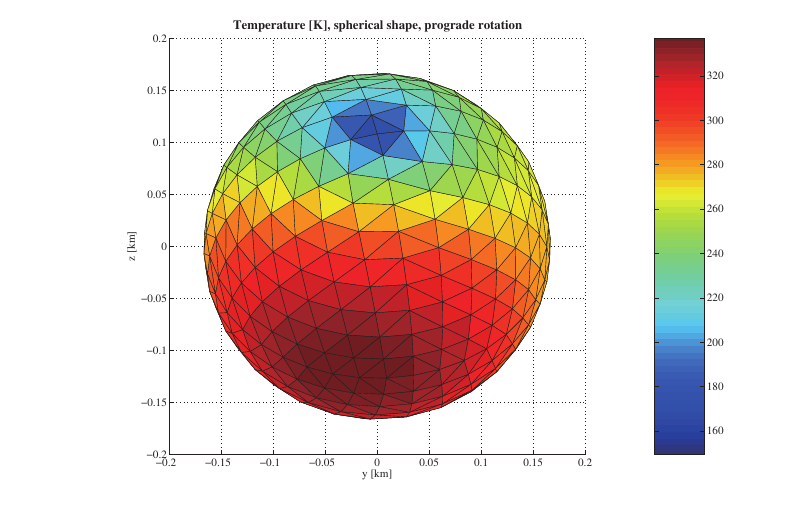}\\
      \includegraphics[width=80mm]{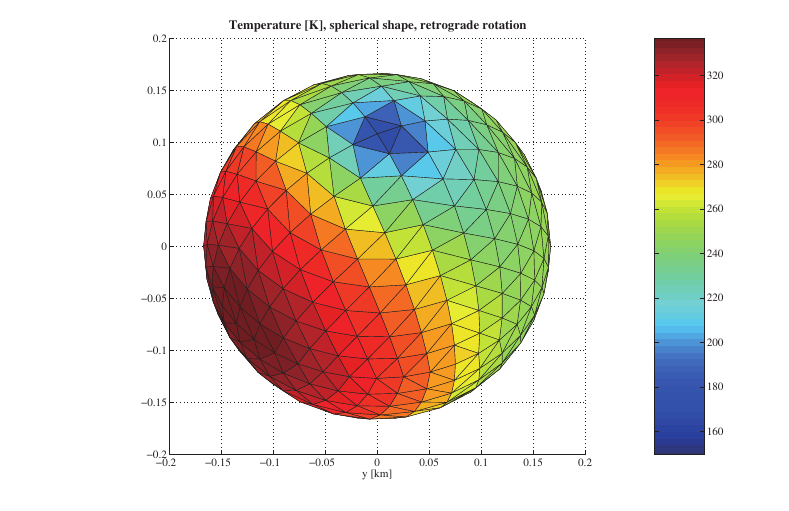}
      \caption{The pro- and retrograde implementation of the spherical
               shape model for the epoch 01-Jul-2004 06:03 UT. For the
               temperature calculation a thermal inertia
               of 700\,J\,m$^{-2}$\,s$^{-0.5}$\,K$^{-1}$ has been used. The
               viewing geometry is in the ecliptic coordinate system (spin vector
               perpendicular to the ecliptic plane) as seen from Earth, projected
               on the sky. The Sun is at a phase angle of 54$^{\circ}$. The asteroid's
               apparent position (Earth-centred) was 312$^{\circ}$ ecliptic
               longitude and -48$^{\circ}$ ecliptic latitude.
       \label{fig:sphere}}
    \end{center}
  \end{figure}

  The shapes and spin-axis orientations are not known for most of the asteroids. It is therefore 
  very instructive to start the radiometric technique with the simplest
  shape model to evaluate the possibilities and limitations of such
  a simple approach.
  In a first attempt to interprete the thermal observations we use
  a spherical shape model with a range of pro- and retrograde spin-vector
  orientations ($\beta_{ecl}^{SV}$\,=\,$\pm$\,30/60/90$^{\circ}$, arbitrary $\lambda_{ecl}^{SV}$)
  and the values specified in table~\ref{tbl:tpm_input}.
  Figure~\ref{fig:sphere} shows the implementation of this model
  for the observation Nr.\ 12 (table~1 in M05) at
  a phase angle of 54$^{\circ}$ (01-Jul-2004 06:03 UT) for a $+90^{\circ}$ prograde (top) and
  $-90^{\circ}$ retrograde (bottom) sense of rotation. The temperature pictures, as seen from
  the observer, are very different and consequently also the connected disk-integrated
  thermal fluxes. The illustration shows that the combination of thermal observations
  from before and after opposition (with either warm or cold terminator) 
  can indicate the true sense of rotation.

  \begin{figure}
    \begin{center}
      \includegraphics[width=80mm]{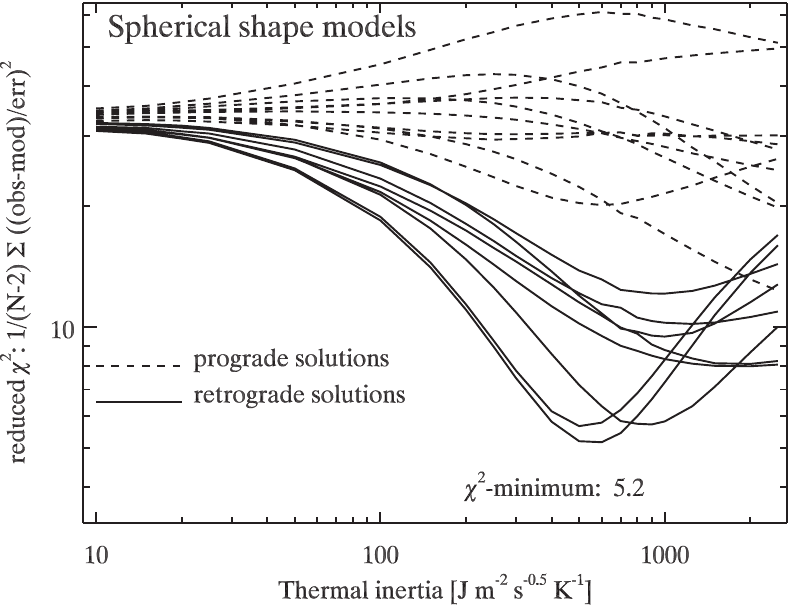}
      \caption{The thermal inertia $\chi^2$ optimisation process
               for all thermal observations, assuming a
               spherical shape model and 18 different pro- and retrograde spin-axes
               orientations (at latitudes $\pm$30$^{\circ}$, $\pm$60$^{\circ}$, $\pm$90$^{\circ}$,
               and arbitrary longitudes).
       \label{fig:ito_sphere}}
    \end{center}
  \end{figure}

  The $\chi^2$-figure (figure~\ref{fig:ito_sphere}) shows that the best agreement
  between observations and model predictions are found for models with a retrograde sense
  of rotation. The prograde rotation options have much higher $\chi^2$-values
  and often show no clear minimum within the very large range of thermal inertias.
  In addition to clear indications for the sense of rotation the $\chi^2$-analysis
  also points towards thermal inertias in the approximate range
  400-1200\,J\,m$^{-2}$\,s$^{-0.5}$\,K$^{-1}$ where the lowest $\chi^2$-values are found.
  The 3 curves with lowest $\chi^2$-minima are connected to
  ($\lambda_{ecl}^{SV}$, $\beta_{ecl}^{SV}$) = (90$^{\circ}$, -60$^{\circ}$),
  (60$^{\circ}$, -30$^{\circ}$), (0$^{\circ}$, -90$^{\circ}$). On basis of our photometric
  data set and without using additional visual light-curve information, it is apparently not
  possible to constrain the spin-axis orientation further within the retrograde domain.

  The connected effective radiometric diameter ($\beta_{ecl}^{SV}\,=\,-90^{\circ}$)
  is 0.31\,$\pm$\,0.04\,km, the geometric albedo 0.30\,$\pm$\,0.06
  (mean and r.m.s.\ of the radiometric solution for the 30 individual observations).
  The diameter/albedo values for the low $\chi^2$-minima connected to the pole-solutions
  (90$^{\circ}$, -60$^{\circ}$) and (60$^{\circ}$, -30$^{\circ}$) are within this
  error range.

\subsection{Using a shape model from light-curve inversion technique}
\label{sec:trishape}

  \citet{kaasalainen05} published a shape model
  with a spin-vector solution based on a large set of remote, disk-integrated
  photometric light-curve observations during the years 2000 to 2004.
  The long time-line allowed to determine an accurate rotation
  period, a high quality pole solution, and a shape estimate.
  The shapes derived from light-curve inversion techniques are
  reproducing the existing set of visual light-curves, but they do not have 
  an absolute size information connected to it.
  The Kaasalainen shape model for (25143)~Itokawa has 1022 vertices and 2040 facets
  and it agreed well with the radar-based solution
  (\cite{ostro05}).

  \begin{figure}
    \begin{center}
      \includegraphics[width=80mm]{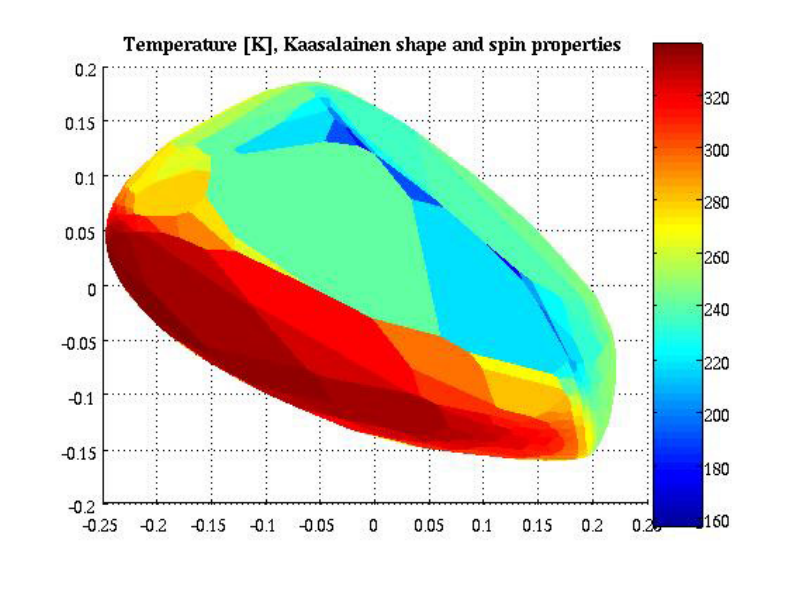}\\
      \includegraphics[width=80mm]{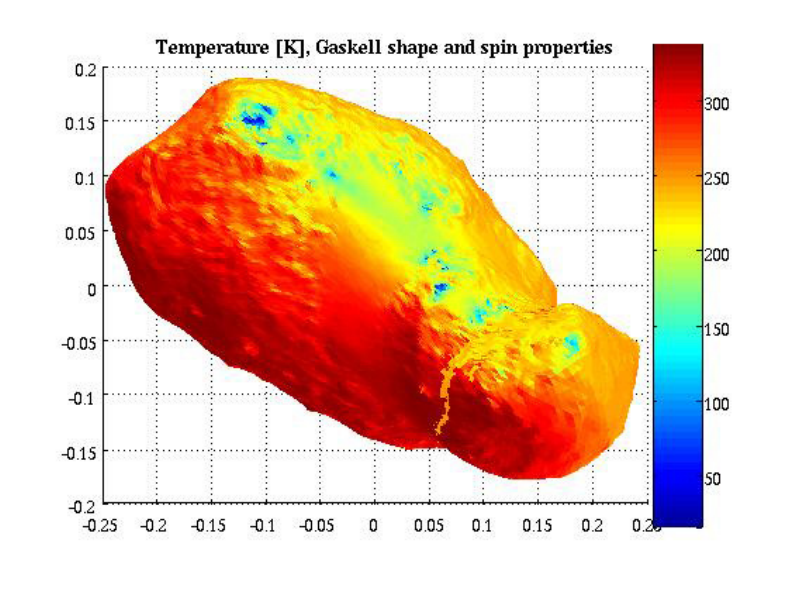}
      \caption{The implementation of the Kaasalainen shape model with 2\,040 facets (top)
               and the Gaskell shape model with 49\,152 facets (bottom)
               for the epoch 01-Jul-2004 06:03 UT. For the
               temperature calculation a thermal inertia
               of 700\,J\,m$^{-2}$\,s$^{-0.5}$\,K$^{-1}$ has been used. The
               viewing geometry is in the ecliptic coordinate system
               as seen from Earth, projected on the sky. The Sun is at a
               phase angle of 54$^{\circ}$. The asteroid's
               apparent position (Earth-centred) was 312$^{\circ}$ ecliptic
               longitude and -48$^{\circ}$ ecliptic latitude.
       \label{fig:trishape_gaskellshape}}
    \end{center}
  \end{figure}

  The rotation parameters are $\beta_{pole}=-89^\circ\pm 5^\circ$ (retrograde
  sense of rotation), $\lambda_{pole}=330^\circ$ for the ecliptic latitude
  and longitude of the pole, and $P=12.13237\pm 0.00008$\,h for the
  sidereal period.
  The zero rotational phase ($\gamma_0 = 0.0^{\circ}$) of this shape
  model is connected to a zero time of T$_0 = 2451933.95456$. Our TPM
  implementation of this Kaasalainen shape model was tested and verified on absolute
  times and rotational phases against the highest quality visual 
  light-curves.
  Figure~\ref{fig:trishape_gaskellshape} (top) shows the temperature-coded (based on our best
  TPM parameters) shape model as seen from the observer on 01-Jul-2004 06:03 UT
  (observational data point number 12 in table~1 in M05).
  
  \begin{figure}
    \begin{center}
      \includegraphics[width=80mm]{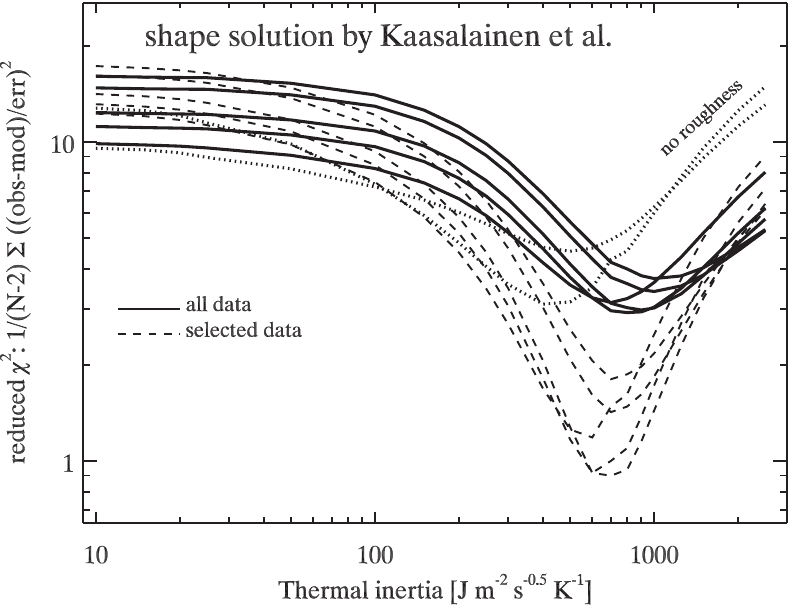}
      \caption{The thermal inertia $\chi^2$-optimisation process
               for all thermal observations (solid lines) and a selected subset
               (dashed lines) assuming the shape model from \citet{kaasalainen05}.
               The two dotted lines represent the corresponding results for a smooth surface
               without additional roughness. The 5 solid lines and the 5 dashed lines
               are the results for different levels of roughness ranging from a
               relatively smooth surface ($\rho$=0.4, f=0.4) to an extremely rough
               surface completely covered by hemispherical craters ($\rho$=1.0, f=1.0).
       \label{fig:ito_trishape}}
    \end{center}
  \end{figure}

  We repeated the $\chi^2$-procedure (see M05 and Sect.~\ref{sec:sphere})
  with the given Kaasalainen spin-vector for a range of thermal inertias.
  The primary goal was to determine the effective size of the scale-free
  Kaasalainen shape-model and to narrow down the possible thermal inertia
  of Itokawa's surface.
  This time we also modified the TPM surface roughness 
  to investigate the influence in the optimisation process (figure~\ref{fig:ito_trishape}). 
  The new minimum of 2.9 in the $\chi^2$-calculations (lowest solid line in
  figure~\ref{fig:ito_trishape}) is now significantly lower than in the case
  of a spherical shape model, indicating that the effects of the non-spherical
  shape are clearly dominating the $\chi^2$-optimisation. But the $\chi^2$-values 
  are still relatively high and show that some data points are not well matched.
  In a second round of $\chi^2$-calculations we deselected all data which
  are marked as taken under bad weather conditions (labeled with $\star$ in M05
  and airmass $> 2.0$ data by \cite{mueller05a}). The 
  results are shown as dashed lines in figure~\ref{fig:ito_trishape}. Now,
  the reduced-$\chi^2$ minimum is at 0.9 and we obtained an excellent match
  between observed fluxes and model predictions.
  Our findings can be summarised as follows:
  (1) Additional surface roughness is needed in the modelling to obtain acceptable
      $\chi^2$-solutions. Without roughness (dotted lines in figure~\ref{fig:ito_trishape})
      the $\chi^2$-minima are a factor 2-3 higher, mainly caused by the poor match to the
      data at shortest wavelengths and smallest phase angles.
  (2) The surface roughness
      influences the calculations. It plays an important role at mid-IR wavelength
      where the thermal infrared emission peaks (see also \cite{mueller02a}):
      roughness variations cause a change in SED slopes and in second order also
      a change in absolute thermal fluxes, with different impact at different phase
      angles. Depending on the surface roughness, the minima
      in figure~\ref{fig:ito_trishape} are shifting slightly and therefore add to the
      thermal inertia uncertainty.
  (3) At a certain level of surface roughness it is not possible anymore to
      distinguish between roughness influence and thermal inertia influence.
      More observational data closer to opposition would be needed to disentangle
      these competing surface properties in the TPM.
  (4) The full dataset and also the selected high-quality dataset show
      the $\chi^2$-minima at similar thermal inertias: The most likely thermal inertia
      solutions are in the range 600-1100\,J\,m$^{-2}$\,s$^{-0.5}$\,K$^{-1}$
      (higher values are connected to higher levels of surface roughness and vice versa).
  (5) The quality limitations of our set of thermal observations determine
      the $\chi^2$-minima. A few low quality photometric observations dominate the
      final uncertainties.
  (6) The corresponding effective radiometric diameter is 0.320$\pm$0.029\,km, the
      geometric albedo 0.299$\pm$0.043 at a thermal inertia of 800\,J\,m$^{-2}$\,s$^{-0.5}$\,K$^{-1}$.
      These values are in excellent agreement with the in-situ results. 
  
%
%
%

\subsection{Using the in-situ shape model}
\label{sec:insitu}
  
\begin{table*}[h!tb]
{\footnotesize
    \caption{Summary of the critical radiometrical properties from different sources. The numbers in
             bold face indicate the current best values.}
      \label{tbl:sum}
  \begin{center}
    \begin{tabular}{lcccl}
      \hline \hline
      \noalign{\smallskip}
             & Size/Shape     & colours \&   & thermal params $\eta$ or                             & Remarks/ \\
      Source & D$_{eff}$ [m]  & geom.\ albedo p$_V$ & $\Gamma$ [Jm$^{-2}$s$^{-0.5}$K$^{-1}$]       & Comments \\
      \noalign{\smallskip}
      \hline
      \noalign{\smallskip}
      \cite{ostro01}                 & 630($\pm$60)$\times$250($\pm$30)\,m & --- & --- & radar 2001 \\
      \cite{sekiguchi03}             & 352$^{+28}_{-32}$\,m & 0.23$^{+0.07}_{-0.05}$ & NEATM\footnotemark[$^{1}$]-$\eta$ = 1.2 & single N-band \\
      \cite{ohba03}                  & a/b=2.1, b/c=1.7; triaxial ellipsoid & --- & --- &  light-curve inversion \\
      \cite{ishiguro03}              & 620($\pm$140)$\times$280($\pm$60)$\times$160($\pm$30)\,m & 0.35$\pm$0.11 & FBM\footnotemark[$^{2}$]-$\eta$ = 1.1 & M$^{\prime}$ \& N-band \\
      \cite{kaasalainen03}           & a/b=2.0, b/c=1.3; triaxial shape & no variegation & --- & light-curve inversion \\
      \cite{ostro04}                 & 548$\times$312$\times$276\,m; 358 ($\pm$10\%)\,m & no variegation & --- & radar 2001 \\
      \cite{ostro05}                 & 594$\times$320$\times$288\,m; 364 ($\pm$10\%)\,m & --- & --- & radar 2001-2004 \\
      \cite{kaasalainen05}           & improved triaxial shape & --- & --- & light-curve inversion \\
      \cite{mueller05a}              & D$_{eff}$=280\,m & --- & $\Gamma$=350 & multi-epoch M$^{\prime}$-/N-data \\
      \cite{mueller05b}              & D$_{eff}$=320$\pm$30\,m & 0.19$^{+0.11}_{-0.03}$ & $\Gamma$=750$\pm$250 & multi-epoch N-/Q-data \\
      \cite{lowry05}                 & a/b$>$2.14 & colours; spec.\ slope & --- & BVRI photometry \\
      \cite{demura06}                & 535$\times$294$\times$209($\pm$1); {\bf 327.5$\pm$5.5\,m} & --- & --- & in-situ \\
      \cite{thomas08}                & a/b=1.9$\pm$0.1 & 0.23$\pm$0.02; H$_V$ & --- & V \& NIR observations \\
      \cite{gaskell08}               & D$_{\mathrm{eff}}$ = 334\,m; {\bf highres.\ shape models} & --- & --- & Hayabusa/AMICA \\
      \cite{bernardi09}              & --- & {\bf H$_V$ \& G-slope}  & --- &  V-band data \\
      \noalign{\smallskip}
      \hline
      \noalign{\smallskip}
      this work (Sect.\ \ref{sec:sphere})   & 310 $\pm$ 40\,m; spin-axis estimate & 0.30 $\pm$ 0.06 & $\Gamma$= 400-1200 & spherical shape \\
      this work (Sect.\ \ref{sec:trishape}) & 320 $\pm$ 29\,m & 0.299 $\pm$ 0.043 & $\Gamma$ = 600-1100 & lc-inversion shape \\
      this work (Sect.\ \ref{sec:insitu})   & --- & {\bf 0.29 $\pm $0.02} & {\bf $\Gamma$ = 500-900} & in-situ shape \\
      \noalign{\smallskip}
      \hline
      \noalign{\smallskip}
      \multicolumn{5}{l}{\hbox{\parbox{160mm}{\footnotesize
          \par\noindent
              {Notes.
               \footnotemark[$^{1}$]{NEATM: Near-Earth Asteroid Thermal Model;}
               \footnotemark[$^{2}$]{FBM: free beaming parameter thermal model.}}}}}
  \end{tabular}
  \end{center}
}
\end{table*}

  One result from the Hayabusa-mission is the highly accurate physical
  description of (25143)~Itokawa. These shape models were produced
  by Robert Gaskell and available in different resolutions from
  {\tt http://sbn.psi.edu/pds/resource/itokawashape.html}.
  The four resolutions correspond to 6$(Q+1)^2$ vertices and 12$Q^2$ facets,
  with $Q=64, 128, 256, 512$.
  The models include many small-scale features on the
  surface seen by the Hayabusa-mission during close flybys.
  The Gaskell shape model is scaled to the true object size
  which corresponds to an effective diameter D$_{\mathrm{eff}}$ = 0.334\,km
  of an equal volume sphere.

\subsubsection{Determination of the geometric albedo}

  The true-size shape model allows now to investigate the influence
  of the aspect angle (defined such that it equals 0$^{\circ}$ when
  observing the North pole and it equals 180$^{\circ}$ when observing
  the South pole of the object) on H-mag calculations and consequently
  also on the determination of the geometric albedo (see also 
  \cite{orourke12} on similar considerations 
  for (21)~Lutetia).

  \citet{bernardi09} measured
  V-magnitudes of Itokawa for a wide range of phase angles and
  determined the corresponding mean light-curve values via a fit
  of synthetic light-curves (using the Gaskell shape model) to the
  observed incomplete light-curves. But depending on the aspect angle,
  the light-curved averaged cross-section can vary between about 320\,m
  and 410\,m! The relevant cross-sections (phase angles $< 40^{\circ}$)
  for the H-mag determination in \citet{bernardi09}
  were all very small (324$\pm$4\,m instead of 332\,m used by \cite{bernardi09}).
  Their derived H-mag of 19.40$^{+0.10}_{-0.09}$\,mag is therefore
  only applicable for this smaller cross-section. The corresponding
  geometric albedo p$_V$ (relations in \cite{bowell89})
  is 0.31$\pm$0.03.

  \citet{thomas08} published a H$_V$ of 19.472$\pm$0.006\,mag
  based on observations taken in 2004 covering a wide phase angle range.
  We calculated the corresponding aspect angles and found again that the
  light-curve-averaged cross-section diameters (at crucial phase angles $< 40^{\circ}$)
  were slightly smaller (just around 0.330\,km) than the effective diameter
  of the Gaskell shape model (of an equal volume sphere). 
  The corresponding geometric albedo p$_V$ is 0.28$\pm$0.02.

  These observations from 2000, 2001 and 2004 cover a huge phase angle range
  between 4 and 129$^{\circ}$ and aspect angles from 50 to about 150$^{\circ}$.
  Merging both datasets and considering the aspect angle limitations of the
  individual sets, we assign a geometric albedo of 0.29$\pm$0.02 for our
  TPM calculations (based on a solar constant\footnote{Active Cavity Radiometer Irradiance
  Monitor (ACRIM) total solar irradiance monitoring 1978 to present (Satellite observations
  of total solar irradiance); access date 2014-01-28; \tt http://www.acrim.com/}
  of 1366\,W\,m$^{-2}$).
  The true, object-connected H-mag, averaged over all aspect angles and
  connected to the average size of an equal volume sphere, is then H$_V$=19.4$\pm$0.1\,mag.
 
\subsubsection{Verification of radiometric size and thermal inertia}

  Similar to our analysis in Sect~\ref{sec:sphere} and \ref{sec:trishape}
  we determined the size, albedo and thermal inertia information through
  our TPM implementation and using the ``Gaskell'' shape model with
  the absolute size as a free parameter. The corresponding $\chi^2$-picture
  is very similar to figure~\ref{fig:ito_trishape}. We obtained minimum
  (reduced) $\chi^2$-values around 3 for the full data-set and just below
  1 for the high-quality sub-set of observations.
  The best match between
  observations and model predictions was found for an effective size of
  0.332$\pm$0.033\,km and a geometric albedo of 0.30$\pm$0.05 (smaller
  errors if we only use the high-quality subset of data). These means
  and standard deviations are connected to a thermal inertia of
  700\,J\,m$^{-2}$\,s$^{-0.5}$\,K$^{-1}$ at the $\chi^2$-minimum and
  assuming an intermediate level of surface roughness. The radiometrical size
  derived in this way is in excellent agreement with the true in-situ size.
  This validates our model implementation and analysis technique.
  But similar to figure~\ref{fig:ito_trishape},
  we see a small shift in thermal inertia when we modify the roughness level.
  An extremely rough surface (f=1.0, $\rho$=1.0: 100\% of the surface covered
  by hemispherical craters with a surface slope r.m.s.\ of 1)
  requires slightly higher thermal inertias
  up to about 900\,J\,m$^{-2}$\,s$^{-0.5}$\,K$^{-1}$.
  A smoother surface (f=0.4, $\rho$=0.4: 40\% of the surface covered
  by shallow craters with a surface slope r.m.s.\ of 0.4) has
  to be combined with a lower thermal inertia
  (500\,J\,m$^{-2}$\,s$^{-0.5}$\,K$^{-1}$) to obtain a good match between
  model predictions and observed fluxes. This possible range for the thermal
  inertia can be translated into a mean grain radius of the surface regolith of
  about 21$^{+3}_{-14}$\,mm (\cite{gundlach13}) which is in excellent agreement
  with the in-situ findings (\cite{yano06, kitazato08}). The heat transport
  within the top-surface layers is therefore dominated by radiation effects
  and material properties play a very small role for the thermal behaviour
  of Itokawa.

  We also tested thermal model solutions without adding any additional roughness.
  The $\chi^2$-test produced an acceptable solution ($\chi^2$-minimum only
  about 15\% higher than for the default roughness case), but the corresponding
  radiometric effective size was below 0.3\,km, well outside the possible range.
  This confirms again that a certain roughness level is needed to
  explain the available observations, even in case of highly detailed and
  structured shape models. In fact, the surface features in the Gaskell
  model still belong to the ``global shape''. They are still large in
  comparison to the thermal skin depth scales\footnote{The 49,152 facet
  shape model has an average facet dimension of $\sim$4\,m whilst the highest
  resolution $\sim$3 million facet shape model has an average facet dimension of $\sim$0.5\,m.
  Both of which are much larger than Itokawa's implied thermal skin depth of about 1\,cm.}.
  Therefore it is needed to add an
  artificial roughness to account for the thermal beaming effect (\cite{lagerros98b}).
  This effect occurs mainly at centimetre scales, with small contributions
  coming from surface porosity on smaller scales (\cite{hapke96};
  \cite{lagerros98a}). Multiple scattering of radiation increases the total amount of
  solar radiation absorbed by the surface and rough surface elements oriented
  towards the Sun become significantly hotter than a flat surface
  (\cite{rozitis11}).
  Without such a ``beaming model'' on top
  of the Gaskell model it was not possible to find a convincing radiometric
  solution for all observations simultaneously. Here we used (as before for
  the spherical and Kaasalainen shape models) the beaming model concept
  developed by \citet{lagerros97}, with $\rho$, the r.m.s.\ of
  the surface slopes and $f$, the fraction of the surface covered
  by craters. The beaming model produces a non-isotropic heat
  radiation, which is noticeable at phase angles close to opposition
  (see also figure~\ref{fig:phase}).
  But it also influences the shape of the SED 
  in the mid-IR which is very relevant for our data set
  (e.g., \cite{mueller02a}).
  But more thermal data closer to
  opposition would be needed to fully characterise surface roughness
  properties of Itokawa.

\subsubsection{Limitations of the TPM}

\begin{figure}
    \begin{center}
      \includegraphics[width=80mm]{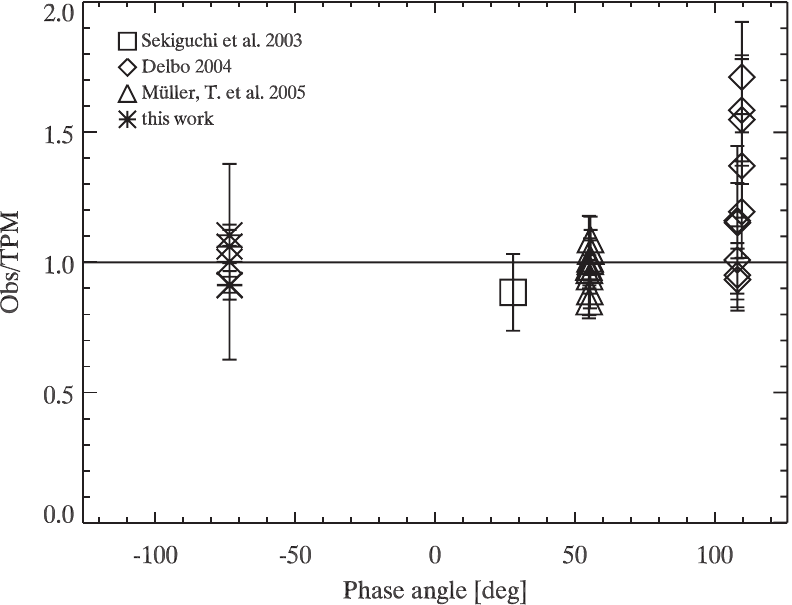}\\
      \includegraphics[width=80mm]{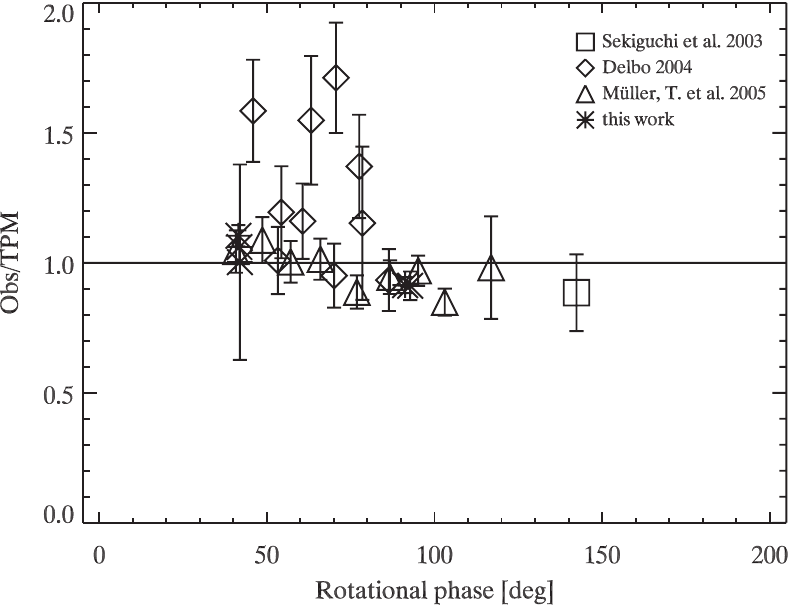}\\
      \includegraphics[width=80mm]{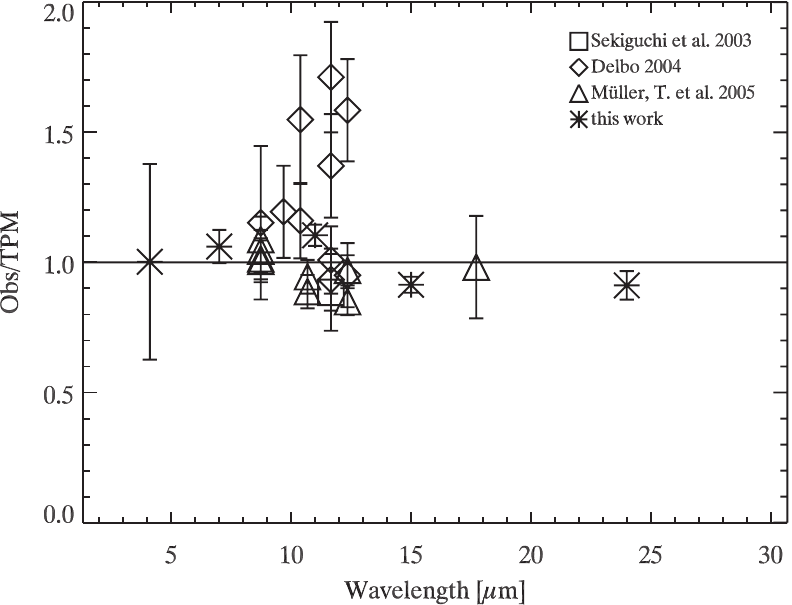}
       \caption{All thermal observations divided by the corresponding TPM prediction
                as a function of phase angle (top), as a function of rotational phase (middle)
                and as a function of wavelength (bottom), with the zero rotational
                phase in the TPM setup defined at JD\,2451934.40110 or 2001-Jan-24 21:37:35 UT).}
       \label{fig:obsmod}
    \end{center}
 \end{figure}

  We combined the Gaskell shape model (accepting the Gaskell size
  scale) now with our derived thermal properties: an intermediate
  roughness level (f=0.7, $\rho$=0.6: 70\% of the surface covered
  by craters with a surface slope r.m.s.\ of 0.6) and a thermal
  inertia of 700\,J\,m$^{-2}$\,s$^{-0.5}$\,K$^{-1}$.
  In figure~\ref{fig:obsmod}
  we present the ratios between observed fluxes and
  the corresponding TPM predictions and show this ratios 
  as a function of phase angle, wavelength, and rotational phase.
  These kind of plots are very sensitive to changes in the thermal properties
  (a full discussion of the influences is given in \cite{mueller02a}): 
  a wrong thermal inertia would lead to slopes in the phase-angle picture
  (top in figure~\ref{fig:obsmod}) with large deviations from 1.0 at the
  largest phase angles, while wrong beaming parameters are dominating at
  the smallest phase angles close to opposition. The wavelength picture
  (bottom in figure~\ref{fig:obsmod}) gives indications about emissivity
  variations and is strongly reacting to beaming parameter variations,
  especially at wavelengths shorter than the peak wavelength.
  The observation/TPM ratios also change with
  different aspect angles and the different sets of measurements are
  not easy to compare. Our optimum TPM solution seems to
  combine the available observations without any obvious remaining trend
  in phase angle, wavelength or rotational phase.

  Nevertheless, there are individual outliers where observations and model predictions
  differ significantly. Some of the data at phase angles above 100$^{\circ}$ 
  (also visible in figure~\ref{fig:obsmod} middle and bottom) are very
  problematic in the TPM calculations. According to \citet{delbo04} the quality
  of some of the measurements is poor, but it could also well be that the TPM
  temperature calculations are not correct and that instead of a 1-D heat conduction
  a 3-D heat conduction would be required for some of the extreme viewing and illumination
  geometries (see also discussions by \cite{davidsson10}). 
  The 1-D heat conduction seems to work fine in case of simple shape models, but
  if a shape model has many small-scale features then the lateral heat conduction might
  influence the true surface temperatures considerably at least at very large
  phase angles.
  Another reason for the offset at large phase angles could be related to the
  approximate treatment of heat conduction inside the craters (see Section~\ref{sec:tpm}).
  \citet{rozitis11} showed that the rough surface thermal emission is enhanced slightly
  on the night side because of the mutual self-heating of the interfacing rough surface
  elements. This could be another explanation why our TPM predictions are lower than
  the observed fluxes at high phase angles in Fig.~\ref{fig:obsmod} (top).

  The $\chi^2$-analysis using the Gaskell shape model
  is almost identical to the analysis using the Kaasalainen shape model:
  We obtained very similar $\chi^2$-values and also the derived thermal
  properties agree very well.
  This shows that our analysis is limited by the number and quality of
  the thermal observation and not by shape information.
  The uncertainties are pure r.m.s.-values
  from the 25 individually derived radiometric solutions, but they 
  reflect to a certain extent also uncertainties in surface roughness
  and thermal inertia and the quality of the mid-IR photometric data points.

\section{Model Comparison and Predictions}
\label{cha:predictions}

\subsection{Thermal light-curve comparison}
\label{sec:lc}

\begin{figure}
    \begin{center}
      \includegraphics[width=80mm]{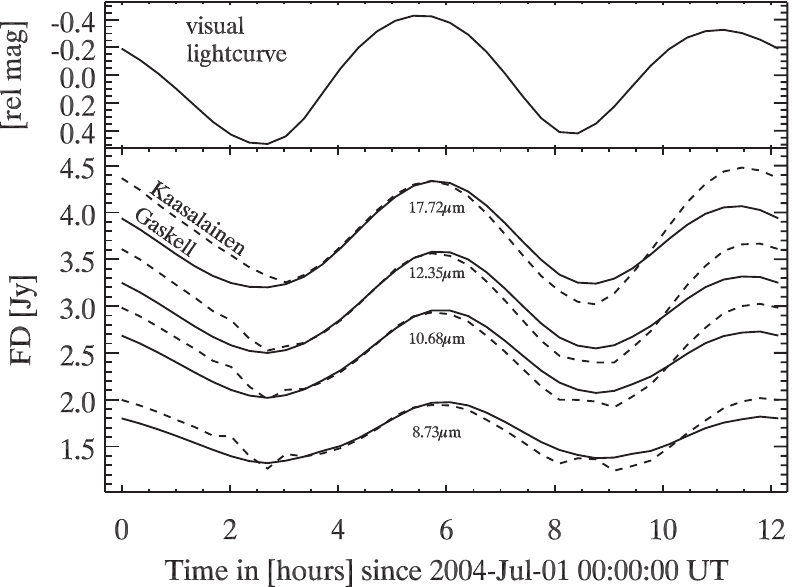}
      \caption{{\bf Top:} The visual light-curve in relative magnitudes, based
               on the Gaskell shape-model. {\bf Bottom:}
               A comparison of thermal light-curves produced
               with the Kaasalainen shape model (dashed lines)
               and the Gaskell shape model (solid lines)
               for the epoch 01-Jul-2004 when the asteroid was
               see at phase angles of 54-55$^{\circ}$.
       \label{fig:tlc}}
    \end{center}
 \end{figure}

Now, with the true shape model at hand, it is interesting
to compare thermal light-curves from the Kaasalainen shape
model with the light-curves from the Gaskell shape model.
In figure~\ref{fig:tlc} we calculated for one full rotation
period with a starting time 01-Jul-2004 at 00:00 UT the visual
light-curve (in relative magnitudes), based on the Gaskell
shape-model (top), and the thermal light-curves (bottom)
at different wavelengths (8.73, 10.68, 12.35, 17.72\,$\mu$m).
At visible wavelength the Kaasalainen-shape model matches very well
the existing light-curves (\cite{kaasalainen05, durech08}). 
In the mid-IR thermal range the
results of the Kaasalainen shape model follows for large fractions
of the light-curve from the Gaskell model. But at specific rotational
phases the differences are significant. The Kaasalainen shape model
produces a larger peak-to-peak thermal light-curve amplitude,
nicely visible in the Q-band at 17.72\,$\mu$m curve
and it also produces bumps and sharp edges at certain phases. 

The relatively large facets in the Kaasalainen shape model
cause artificial structures in the predicted thermal light-curve
while the Gaskell model produces astonishingly smooth and regular curves.
In the given observing set there are also severe 
differences in the peak-to-peak variations from both models
(about 10-20\%, depending on the wavelength), but the
deviations vary not only with wavelength, but also with aspect angle.
The minima in the thermal light-curve are also broader than for the
visual light-curve. This is an effect of the object's
shape at a given rotation angle combined with the high thermal
inertia which smoothes out the rapid changes of illuminated
surface areas.
It can be concluded that the Kaasalainen shape model, although it was
derived from light-curve observations and matches nicely visual
light-curves for a wide range of observing geometries, still
has shortcomings in the context of thermal light-curves.
Or in other words: the light-curve inversion technique might
benefit from using thermal light-curves
and the resulting shape models and spin-axis orientations
would come closer to reality.

\subsection{TPM predictions using the Gaskell shape model}

The Gaskell shape model in combination with the derived 
and validated thermophysical properties allows now to
do more generalised studies. What can be learnt from thermal spectra or
light-curve measurements at different wavelengths? How does
the opposition effect look like at thermal wavelengths?
What are the key observing geometries for successful
radiometric calculations?

 \begin{figure}
    \begin{center}
      \includegraphics[width=80mm]{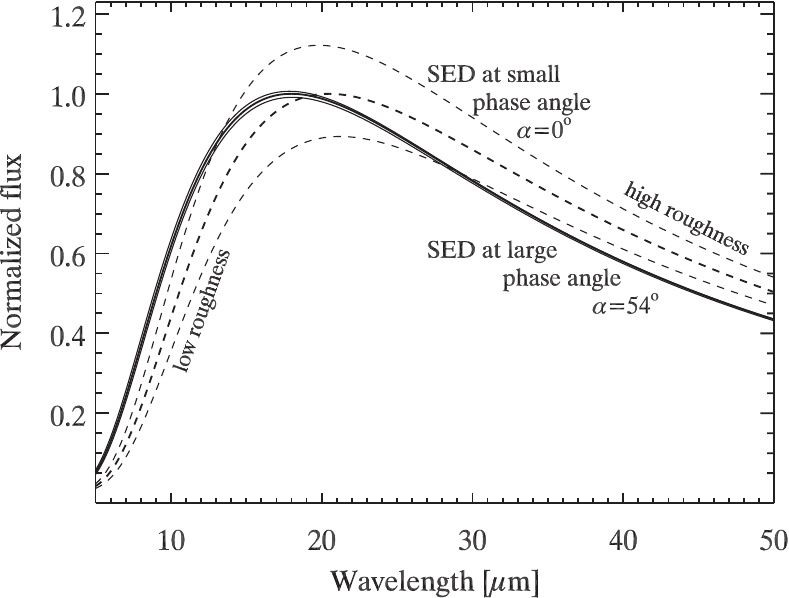}\\
      \includegraphics[width=80mm]{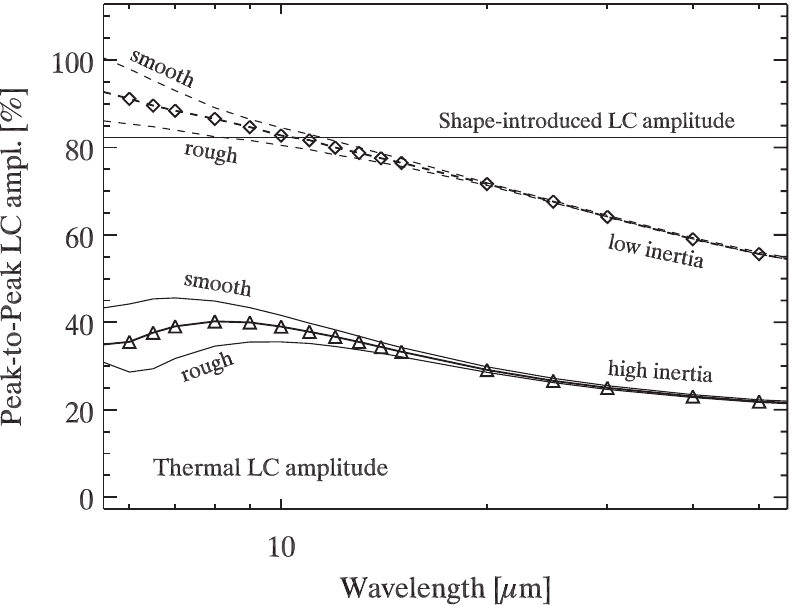}
      \caption{The Gaskell shape model combined with thermal properties: Thermal
               effects as a function of wavelength. {\bf Top:} The normalised
               SEDs for observations close to opposition
               (dashed lines) and at 54$^{\circ}$ phase angle (solid lines)
               for a thermal inertia of $\Gamma$=1000\,J\,m$^{-2}$\,s$^{-0.5}$\,K$^{-1}$.
               The strong roughness influence at small phase angles is clearly
               visible (thermal opposition effect).
               {\bf Bottom:} The thermal light-curve (LC) amplitudes (peak-to-peak in
               [\%] of the absolute thermal flux) for the observing constellation
               on 01/Jul/2004 (54$^{\circ}$ phase angle).
               For the ``low thermal inertia'' case we
               used 15\,J\,m$^{-2}$\,s$^{-0.5}$\,K$^{-1}$ (typical value
               for large main-belt asteroids; \cite{mueller99})
               and for the ``high thermal inertia''
               1000\,J\,m$^{-2}$\,s$^{-0.5}$\,K$^{-1}$. The pure
               shape-caused light-curve amplitude (as seen in visual
               light) is indicated by the horizontal line.
               The peak-to-peak amplitude decreases significantly 
               for longer wavelengths and for higher thermal inertias.
       \label{fig:tlcamp}}
    \end{center}
  \end{figure}

Figure~\ref{fig:tlcamp} (top) shows model predictions (now only using the
Gaskell shape model) for a wavelength range from 5 to 50\,$\mu$m.
Both SEDs at intermediate roughness level were normalised to 1.0
at the thermal emission peak wavelengths. Nevertheless, one can
see that the SED at small phase angles (dashed lines) are higher
(in the longer wavelength range) than the SEDs at larger phase angle
(solid lines). The surface roughness and its thermal-infrared beaming effect
enhance the observed thermal emission at low phase angles. This is 
then balanced out by a reduced thermal emission at larger phase angles
to conserve energy (see also \cite{mueller02a}).
The SEDs close to opposition (dashed lines) are strongly influenced
by variations in the roughness, while at larger phase angles 
(solid lines) the roughness properties play only a minor role.
The flux enhancements observed at low phase angle are dominated by
limb surface enhancements. Some of the surface elements inside craters
located near the terminator are orientated towards the Sun.
The corresponding temperature enhancements are much greater than
those achieved at the bottom of craters near the subsolar region.
The resulting beaming effect is more efficient at low phase angles,
where the observer is able to see the illuminated crater walls which
produce additional thermal flux. This limb-brightening effect has
been seen in spatially resolved measurements, and it has been
successfully modelled for the Moon (\cite{rozitis11}) and
for (21)~ Lutetia (\cite{keihm12}).
The lack of small phase angle thermal observations explains our
difficulties to find a robust solution for Itokawa's
surface roughness. Observations close to opposition would constrain these
properties much better. Figure~\ref{fig:tlcamp} (bottom) shows the behaviour
of the thermal light-curve (at phase angle 54$^{\circ}$) as
a function of wavelength for two values of the thermal inertia.
The ``low thermal inertia'' case represents
typical main-belt values (\cite{mueller99})
caused by a very well insulating dust regolith on the surface
(15\,J\,m$^{-2}$\,s$^{-0.5}$\,K$^{-1}$). For comparison, the Moon has
a thermal inertia of 39\,J\,m$^{-2}$\,s$^{-0.5}$\,K$^{-1}$
(\cite{keihm84}).
The ``high thermal inertia'' case corresponds to our best solution for
the thermal properties of (25143)~Itokawa. In the given geometry
the peak-to-peak brightness variation during one full rotational
period is changing dramatically with wavelength. At mid-IR
the amplitude in the ``low thermal inertia'' case can even exceed the
pure shape-caused brightness variation! The thermal light-curve
amplitude decreases by more than a factor of 2 from mid-IR to
far-IR wavelengths and this effect is almost independent of phase
angle. In the high thermal inertia case (Itokawa) the overall values are
significantly smaller (much smaller than the shape-introduced amplitude)
and also the change with wavelength is smaller. Close to opposition the
light-curve amplitude behaviour in the high thermal inertia case is more complex
and does not show a clear trend with wavelength anymore.
Overall, the peak-to-peak thermal light-curve amplitude decreases 
for higher thermal inertias and at longer wavelengths. Uncertainties in
the light-curve amplitudes due to stronger influences of surface roughness
increase for observations close to opposition.

 \begin{figure}
    \begin{center}
      \includegraphics[width=80mm]{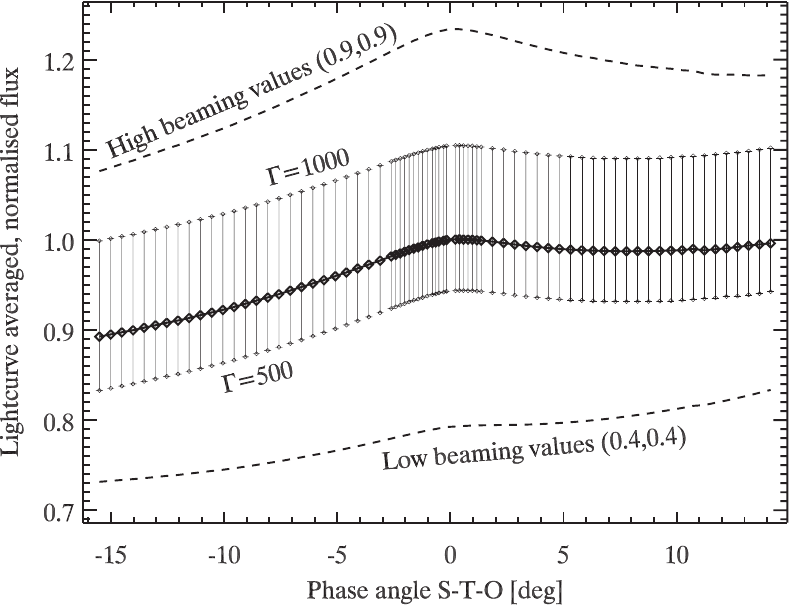}\\
      \includegraphics[width=80mm]{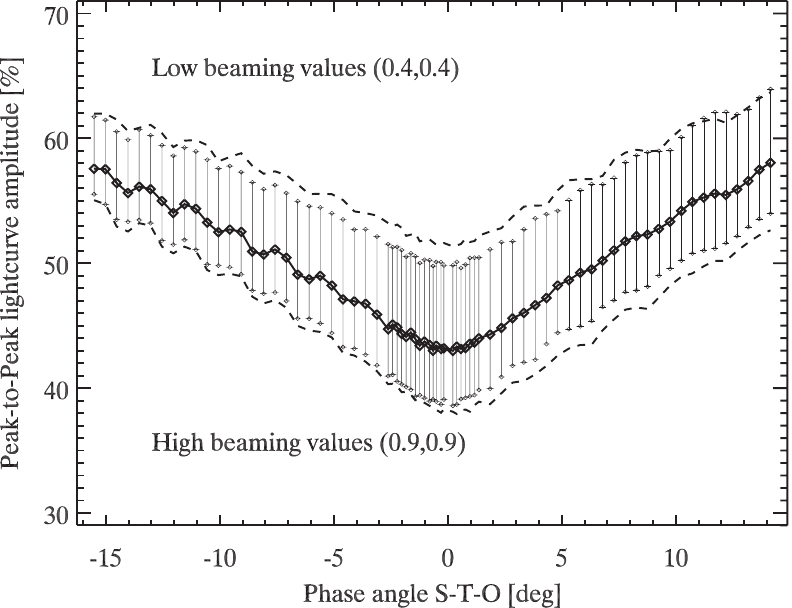}
      \caption{TPM/Gaskell predictions for Itokawa-opposition in 2011.
               The corresponding times are:
                          03-Apr-2011: -15.0$^{\circ}$ (trailing the Sun),
                          13-Jun-2011:   0.1$^{\circ}$,
                          27-Aug-2011: +15.0$^{\circ}$ (leading the Sun).
               {\bf Top:} Normalised (by the asteroid distance to
                       Sun and Earth) 10\,$\mu$m fluxes as a function of
                       phase angle. Uncertainties
                       due to thermal inertia (range from 500-1000\,J\,m$^{-2}$\,s$^{-0.5}$\,K$^{-1}$)
                       are given with error bars, uncertainties due to roughness
                       (range for ($\rho$,f) from (0.4,0.4) to (0.9,0.9)) are
                       indicated by the dashed lines.
                       The opposition ``peak'' at small phase angles
                       can be seen, as well as the influence of surface roughness
                       on the peak-hight.
              {\bf  Bottom:} the peak-to-peak light-curve variation at 10\,$\mu$m in percent
                       as a function of phase angle. The relative light-curve amplitude
                       is significantly smaller close to opposition. The influences
                       of roughness and thermal inertia change slightly with phase angle.
       \label{fig:phase}}
    \end{center}
  \end{figure}

In figure~\ref{fig:phase} we averaged the flux predictions over a complete rotation
and looked at the changes with phase angle covering large periods before and
after opposition. The predictions have been normalised at $\alpha$\,=\,0$^{\circ}$
and for default roughness. The flux change is dominated
by the distance change between Earth and asteroid. The error bars indicate
the uncertainties due to thermal inertia, the dashed lines give the boundaries
for very low and high beaming values, i.e.\ low and high surface roughness.
At small phase angles there is another effect visible:
the thermal opposition or beaming effect. At small phase angles it is possible
to see warmer temperatures inside the small-scale surface structures (modelled
by the crater-like TPM roughness implementation), mainly located near the 
terminator (see above). Overall, this leads to enhanced thermal fluxes
at small phase angles which is not the case in simple thermal models with
constant phase angle corrections or if the surface roughness is omitted.
During both oppositions also the total
peak-to-peak light-curve amplitude changes significantly with the smallest
amplitudes close to opposition.
These figures show that each data set puts different constraints
on the thermal properties.
Observations at small phase angles show the smallest thermal
light-curve amplitudes (bottom of figure~\ref{fig:phase}).
Thermal light-curve measurements at large phase angles ($> 30^{\circ}$)
and close to the emission peak (in the mid-IR for NEAs)
are providing the most stringent constraints on the thermal inertia:
the light-curve amplitude is maximal and the beaming influence is very small.

 \begin{figure}
    \begin{center}
      \rotatebox{90}{\includegraphics[width=80mm]{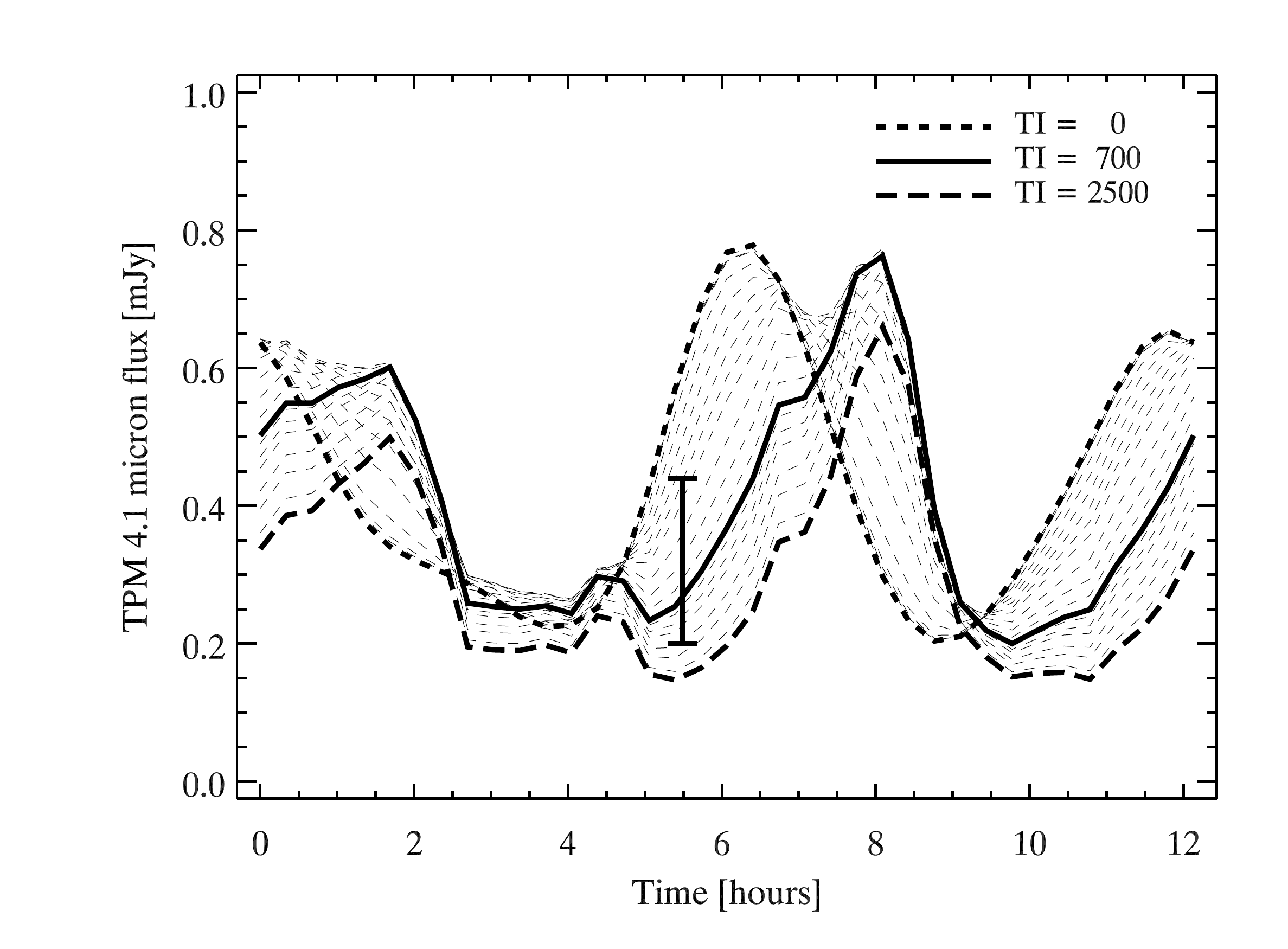}}\\
      \rotatebox{90}{\includegraphics[width=80mm]{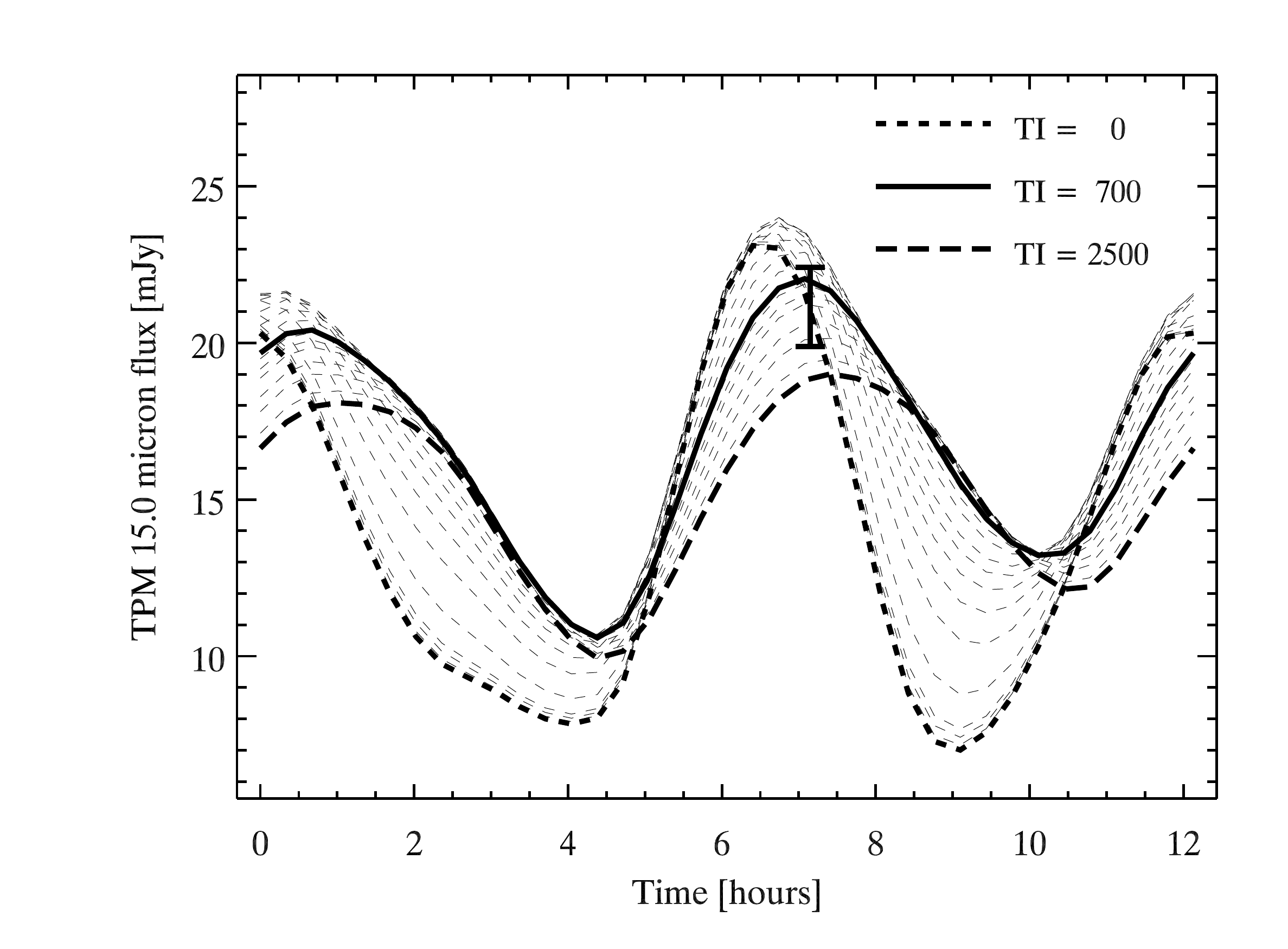}}
      \caption{TPM/Gaskell thermal light-curve predictions for Itokawa during
               the Akari observing epochs for a wide range of thermal inertias.
               {\bf Top:} calculated thermal light-curves at 4.1\,$\mu$m together
                          with the observed Akari data point.
              {\bf  Bottom:} calculated thermal light-curves at 15.0\,$\mu$m
                          together with the observed Akari data point.
       \label{fig:akari_tlc}}
    \end{center}
  \end{figure}

We also looked into synthetic thermal light-curves for the Akari observing epochs
and using the Gaskell shape model. Figure~\ref{fig:akari_tlc} shows these light-curves
at 4.1 (top) and 15.0\,$\mu$m (bottom) for a wide range of thermal inertias. The
TPM code produces absolute fluxes which compare very well with the derived Akari
flux densities which are over-plotted in the figures. Our calculations show that 
the thermal light-curves change dramatically with changing thermal inertia.
At 4.1\,$\mu$m the light curve shape goes from a relatively smooth curve with two
different peak levels (low thermal inertia) to a much more structured curve with
sharp turns (high thermal inertia). The amplitude is not very much affected, but
the delay times of the curves increase with increasing thermal inertia.
In the high thermal inertia case the thermal emission peaks happen much later
and delay times of up to 2 hours are found. At these short wavelengths it is
mainly the hottest sub-solar terrains which dominate the observed fluxes and
flux changes. But due to the roughness-thermal inertia degeneracy (e.g.\
\cite{rozitis11}) the short-wavelength observations are not ideal for
deriving highly reliable size-albedo solutions. However, the shift of the
thermal curves with respect to a purely shape-driven light-curve are
closely connected to the object's thermal inertia. Monitoring light-curve
changes over part of the rotation period is therefore the recommended
approach at the short wavelengths.
At 15.0\,$\mu$m the situation is different: the light-curves are in general
much smoother and their shapes do not change much with thermal inertia,
and the peak thermal emission delay times are smaller than at shorter wavelengths.
These measurements usually lead to more robust size-albedo due to the
reduced influence of the roughness-thermal inertia degeneracy (e.g.\
\cite{mueller02a}). Accessing the object's thermal inertia is more 
difficult: either one has to measure the thermal light-curve amplitude
(which is substantially decreasing with increasing thermal inertia) or
one has to combine the measurements with observations at different
phase angles. It is also worth to note here that the thermal light-curve
shapes, amplitudes, delay times for a given object also depend on the
phase angle. The Akari measurements were taken at a phase angle of 73$^{\circ}$
where the illuminated (hot) part of the surface dominates the short
wavelength observations, while at longer wavelengths there are also
flux contributions from the warm, non-illuminated parts which have just
rotated out of the Sun.


\section{Discussion}
\label{cha:discussion}

\subsection{TPM with spherical shape model}

The interpretation of the 30 thermal infrared observations by
using only a spherical shape model (Sect.~\ref{sec:sphere}) was very
successful: the radiometric effective size prediction lies within a
few percent of the true, in-situ value, the radiometric albedo prediction
agrees within the error-bars with \citet{bernardi09}, 
the retrograde sense of rotation was clearly favoured in the
optimisation process, and thermal inertia values in the range
500-1500\,J\,m$^{-2}$\,s$^{-0.5}$\,K$^{-1}$ are the most likely ones.

But the calculations benefited from the proximity of the assumed
spin axis orientation with the true orientation and from using
a realistic rotation period. If an object's rotation period is
very different from the TPM assumptions and/or the spin vector is
far away from the ecliptic pole direction the TPM predictions will
be less reliable. Also the number of thermal observations, their
distribution in phase angle space, in wavelength space, in rotational
phase space, in aspect angle space and observing geometry influence
the quality of the TPM outcome.

  Overall, the reduced $\chi^2$-values are relatively high, indicating that the
  spherical shape is not always allowing to produce model predictions within
  the given observational errors. The model predictions for the few cases of extreme
  cross-sections dominate the overall $\chi^2$-sums. But here the observations
  are statistically distributed over rotational phases and aspect angles and
  the $\chi^2$-values remain sensitive to basic object properties like thermal
  inertia and sense of rotation. If only very few thermal observations would be available,
  even the distinction between pro- and retrograde solutions could fail for cases
  where shape effects and/or phase angle effects interfere with thermal signatures
  in the terminator region. \citet{mueller02a} demonstrated the 
  capabilities and limitations of the ``radiometric method'' to determine the
  sense of rotation via very spherical shape models. Based on a sample of 9
  main-belt asteroids with comparable sets of thermal observations before and
  after opposition, the thermal data indicated the correct sense of rotation
  for 8 objects. In one case it failed, very likely due to shape and cross-section
  effects and data sets which did not cover the full rotation periods.

Nevertheless, our extremely simplified model approach shows the
large potential of the radiometric technique: Assuming a spherical
shape in combination with a realistic rotation period and a range
of spin-axes
it is possible to derive the sense of rotation and very accurate
values for the effective size.
But the key to robust results is the analysis of
all thermal observations simultaneously, combined with all
available information from visual photometry and light-curves
observations.
The very accurate radiometric diameter prediction has
to be seen in comparison with the radar techniques:
(\cite{ostro04, ostro05})
used delay-Doppler images of (25143)~Itokawa obtained at Arecibo and Goldstone
to predict a size of 594\,$\times$\,320\,$\times$\,288\,m ($\pm$\,10\%), i.e., an
effective diameter of 379.7\,m which is 16\% higher than the 
\citet{fujiwara06} value of 327.5\,m. In addition,
the radar-technique can only be used if the object has a close 
encounter with Earth, while thermal observations are much easier to
obtain.

An important element of the thermal analysis is the reliability of the
H- and G-values. \citet{bernardi09} published new
values based on extensive light-curve data from phase angles close
to opposition out to 115$^{\circ}$.
The new H-magnitude differs by 0.5\,mag from the ones used
in M05. This influences the albedo considerably (see discussion in M05)
while the radiometric diameter is less affected. The difference in
albedo (0.19$^{+0.11}_{-0.03}$ in M05 and 0.299$\pm$0.043 from
Sect.~\ref{sec:insitu}) is mainly caused by this 0.5\,mag change
in absolute magnitude.

\citet{hasegawa08}, and later \citet{mueller11}, applied the TPM technique to
thermal observations of the Hayabusa2 sample return target (162173)~1999\,JU$_{3}$.
First, by using a spherical shape model, then by a modified shape model
which would match the existing visual light-curves in amplitude and
in phase. The quality of the resulting radiometric properties (diameter, albedo, thermal
inertia, ...) suffered from an unknown spin vector orientation at the
time of the calculations. Nevertheless, the example of
(162173)~1999\,JU$_{3}$ shows nicely the potential and the
limitation of applying this method without knowing the precise shape nor
the spin vector orientation.

\subsection{TPM with Kaasalainen-/Radar-shape models}

Using the shape model (together with the spin-vector orientation)
from light-curve inversion techniques led to a radiometric diameter
which was within 2\% of the true in-situ result
(M05; \cite{fujiwara06}). The thermal observations
put very strong constraints on the radiometric diameter.
If, in addition, the H-G values are reliable, i.e., representing
light-curve-averaged properties, or if simultaneous V-magnitudes
are available, then both, the effective diameter and the albedo can
be derived with high accuracy.
Alternatively, the size information from radar techniques or 
occultation measurements can be tested with TPM calculations:
Is it possible to find realistic thermal properties to explain
thermal measurements and the independent size values simultaneously?
M05 tried such an approach with Itokawa's radar size, but they
could not find any acceptable match with the observed fluxes
and they concluded that the radar size must be too large.

Our optimisation process clearly has limitations: Figure~\ref{fig:ito_trishape}
demonstrates that the best solutions from using all 30 data points
were strongly affected by a few less reliable observations.
Thermal observations suffer in some cases from poor absolute
calibration (or badly documented calibration), from high humidity
weather conditions and/or high air-masses, or from uncertainties
in the definition of band-passes with errors in the colour
correction terms.
The plots with observations/model ratios as a function of 
wavelength, phase angle and rotational phase revealed in
many cases the outliers, at least if a high quality shape model 
(covering many aspect angles) is available.

Another limiting factor is the uncertainty in surface roughness.
An artificial roughness model is needed to explain typical mid-IR
asteroid spectra (e.g., \cite{barucci02, mueller04, dotto00, mueller02}), 
but the precise values or the
interpretation of roughness is more complex:
The thermal infrared observations are disk-integrated
observations, averaging over very diverse surface regions. The
resulting disk-integrated thermal properties, especially the
roughness, are therefore of limited significance for characterising
individual regions on the surface. Our ``default'' roughness model
explains the existing observations very well, although it works
in certain observing geometries better than in others, but more observations
close to opposition (see figure~\ref{fig:tlcamp}) or at shorter wavelengths
(e.g.\ \cite{mueller02a} or \cite{rozitis11}) would allow to 
narrow down the range in the possible roughness
parameter space or to even to distinguish between hemispheres or large
surface regions with different roughness properties.
It should also be noted here that the roughness influences
radiometric size and albedo solutions strongly if thermal
infrared observations are only taken close to opposition
and/or at short wavelengths in the Wien-part of the thermal emission.

Overall, the step from spherical to Kaasalainen shape model changed only
slightly the effective diameter, geometric albedo and thermal inertia values,
but the reliability improved significantly: The $\chi^2$-minima shrank from $\sim$5
to about 2.

\subsection{TPM with Hayabusa/Gaskell-shape model}

The availability of the full, high resolution shape model
with spin-vector orientation and absolute size information
(\cite{demura06}) allows to obtain confidence
in the radiometric technique: Using only the shape model and
spin-vector orientation in combination with the set of 
thermal infrared observations we found the
best fit to all data at a thermal inertia of
700\,J\,m$^{-2}$\,s$^{-0.5}$\,K$^{-1}$ 
and under the assumption of a ``default roughness''.
The corresponding radiometric effective diameter agrees
then nicely with the true value. The radiometric diameter
itself has a formal uncertainty of $\pm$8\% when taking
the r.m.s.-residuum from the 25 observations or
$8/\sqrt{25} = \pm$1.6\% when accepting that the observations
represent 25 repeated measurements of the same property, i.e.,
the effective diameter of a equal volume sphere. The radiometrically derived albedo
has a r.m.s.-value of $\pm$15\% or a statistical error of
the mean of about $\pm$3\%. But in case of the albedo the
dominating uncertainty is from the H-magnitude. The $\pm$0.09/0.10\,mag
error in H-value of 19.40\,mag in combination with the true
in-situ size value corresponds to a geometric albedo of
0.29\,$\pm$\,0.03, i.e., $\pm$10\%. The influences of uncertainties
in the thermal properties and the surface roughness only
play a minor role in the final solutions which can be seen
in figure~\ref{fig:ito_trishape} 
where the different curves have a very small scatter close
to the optimum solutions. On the other hand, even the high
resolution shape models require the addition of roughness on the
surface to match the measured fluxes. These high resolution
models include rocks, boulders, craters, valleys and mountains,
i.e., the intermediate scales, but the centimetre scale is
missing. The centrimetre-scale roughness plays an important role, because
that is where the energy is reflected, absorbed, emitted, or
conducted 
in the regolith.
Images from the Hayabusa mission with a spatial resolution
of 70\,cm per pixel (e.g., \cite{saito06})
reveal very diverse surface morphology on intermediate scales
(smooth, rough terrains) and very likely also on micrometre
scale (solid boulders, smooth terrains, cratered rough terrains).


The radiometric technique can be
considered as a very reliable (remote) technique for the determination
of the effective size of small bodies. And the size information
is needed not only for potentially hazardous objects, but
also to investigate densities, size-frequency distributions,
binary/multiple systems, formation process, etc.
The technique only depends
on the availability of thermal infrared observations and is not limited
to close Earth encounters (like the radar-technique) or to only
the very largest main-belt asteroids (like the adaptive optics
technique). 

The thermal inertia is an important parameter for small
bodies in the Solar System, but only known for very few
targets (e.g., \cite{delbo07}). It is relevant
when calculating the main non-gravitational orbit perturbations
over time spans of centuries, caused by the Yarkovsky effect.
It is a result of the way the asteroid's rotation affects the surface
temperature distribution and therefore the anisotropic thermal
re-emission (\cite{vokrouhlicky00}).
The thermal inertia might also help to distinguish between
solid rock surfaces which are expected to have very high thermal
inertias well above 1000\,J\,m$^{-2}$\,s$^{-0.5}$\,K$^{-1}$
and regolith-covered bodies, like the Moon or large
main-belt asteroids. (25143)~Itokawa has an
intermediate thermal inertia, possibly indicating a rubble
pile structure where seismic waves reorganise the body's interior
and the surface frequently and the formation of a thick regolith
(with thermal inertias below 100\,J\,m$^{-2}$\,s$^{-0.5}$\,K$^{-1}$)
is hampered. The reason for having very little fine regolith 
could also be related to the weak gravity on small, low density
objects. Small asteroids tend to have in general higher thermal
inertias (\cite{delbo07}), but there are exceptions
(e.g., \cite{mueller04b}).

\section{Conclusions}
\label{cha:conclusions}

(25143)~Itokawa is an extremely important test case for the 
validation process of thermophysical model techniques. The existing
mid-IR observations allowed to evaluate the possibilities and limitations
of different levels of complexity within our TPM implementation. Even
in cases where very little is known about shape and spin behaviour it
allows to derive reliable properties, like the effective size, the albedo
and thermal properties without using any artificial model fudge factors.
But the outcome is tightly connected to the availability and quality
of thermal infrared observations. Ideally, the observations
should (i) cover a sufficient wavelength range around the emission peak
(several photometric bands or mid-IR spectra);
(ii) include measurements before and after opposition;
(iii) cover a large phase angle range, including measurements close to
opposition; and, 
(iv) include a significant range of rotational phases and/or substantial
parts of the thermal light-curve.
It is also important to note here that some of the mid-IR bands might
be affected by silicate emission (e.g.\ \cite{emery06}), but the effect
on size-albedo solutions can only be estimated if full thermal spectra
are available.

Our analysis showed that for (25143)~Itokawa the interpretation
of surface roughness properties is limited mainly due to the lack 
observations close to opposition. Nevertheless, Itokawa's size,
albedo and thermal inertia have been derived with unprecedented accuracy
by only using remote, disk-integrated observations: a shape model from
standard light-curve inversion technique and thermal infrared observations from
ground and from AKARI. The optimum radiometric size agreed
within 2\% with the true value derived from Hayabusa measurements.

The TPM predictions using the true in-situ shape model showed: 
(i) that the shape models from light-curve inversion techniques
    produce artefacts in thermal light-curves and that low
    thermal inertia object (e.g., large main-belt asteroids) can have
    light-curve amplitudes exceeding the pure shape-introduced values;
(ii) that the SEDs taken close to opposition are strongly influenced by properties
    of the surface roughness (figure~\ref{fig:tlcamp}, top) leading to a strong
    degeneracy between roughness and thermal inertia effects;
(iii) that there exists a thermal opposition effect and how it looks like (figure~\ref{fig:phase}, top), 
(iv) how the thermal light-curve amplitude changes with phase angle (figure~\ref{fig:phase}, bottom); 
(v) that the thermal light-curve amplitudes decrease with wavelength
     and for higher thermal inertias (figure~\ref{fig:tlcamp}, bottom);
(vi) that thermal light-curve delay times increase with thermal inertia and 
     decrease for longer wavelengths (figure~\ref{fig:akari_tlc});
(vii) that the thermal inertia and the sense of rotation play
      a big role when interpreting thermal infrared observations at large phase angles.
(viii) that even the high resolution Gaskell shape model still require
      an additional small-scale roughness to explain the observed
      infrared fluxes.

Our findings are supported by TPM analysis of other spacecraft
target asteroids. In this context one should mention some of the
recent studies on
(4)~Vesta (\cite{mueller98}; \cite{mueller02}; \cite{leyrat12}; \cite{keihm13}),
(21)~Lutetia (\cite{mueller06}; \cite{lamy10}; \cite{orourke12}; \cite{keihm12}),
(433)~Eros (\cite{mueller07}), and
(2867)~Steins (\cite{lamy08}; \cite{groussin11}; \cite{leyrat11}).
In all these cases the radiometrically derived properties are remarkably
consistent with the spacecraft investigations documented by \citet{russell12} for
(4)~Vesta, by \citet{sierks11} for (21)~Lutetia, by \citet{thomas02} for (433)~Eros,
and by \citet{keller10} for (2867)~Steins. The TPM radiometric technique 
is very powerful in deriving highly reliable absolute sizes, albedos, and thermal
inertias from remote disk-integrated thermal measurements. But each of the
available data points has to be considered in its true illumination and observing
geometry. In this way it is also possible to extract information on the object's
spin properties as well as on its shape, especially for objects where standard
lightcurve inversion techniques have difficulties to determine these parameters.

The validated model techniques can easily be used for other targets, 
including near-Earth and main-belt asteroids, trans-Neptunian objects
or inactive cometary nuclei. The plots and figures can also be used
to optimize observing strategies to exploit the full TPM capabilities.
The key ingredient for the full exploitation of thermophysical model
techniques and for the determination of reliable object properties
is the availability of well-selected and well-calibrated thermal infrared
observations covering many aspect angles.

\bigskip

S.~Hasegawa was supported by Space Plasma Laboratory, ISAS, JAXA. 
We would like to thank Johan Lagerros for very useful discussions 
and Robert Gaskell for providing the necessary documentation for
the implementation of Itokawa's shape model.

\end{document}